\documentclass[twocolumn,showpacs,preprintnumbers,amsmath,amssymb]{revtex4}
\usepackage{graphicx}
\usepackage{epsfig}
\newcommand{\Slash}[1]{\ooalign{\hfil/\hfil\crcr$#1$}}
\begin{document}

\preprint{hep-ph/0406055}

\title{
CP asymmetry, branching ratios, and isospin 
breaking effects of $B \to K^*\gamma$ 
with the perturbative QCD approach}

\author{M. Matsumori }%
\email{ mika@eken.phys.nagoya-u.ac.jp}
 \affiliation{
Department of Physics, Graduate School of Science, Nagoya University.
}

\author{A. I. Sanda}%
 \email{sanda@eken.phys.nagoya-u.ac.jp}
 \affiliation{
Department of Physics, Graduate School of Science, Nagoya University.
}

\author{Y.-Y. Keum}%
 \email{yykeum@phys.sinica.edu.tw}
 \affiliation{
Institute of Physics, Academia Sinica, Taipei, Taiwan 115, Republic of
 China.
}

\begin{abstract}
The main contribution to the
radiative ${B\rightarrow K^* \gamma}$ mode is from
penguin operators which are quantum corrections.
Thus, this mode may be useful in the search for physics beyond 
the standard model.
In this paper, we compute the branching ratio, direct CP asymmetry, and 
isospin breaking effects
within the standard model in the framework of perturbative QCD, and 
discuss 
how new physics might show up
in this decay.
\end{abstract}
\maketitle

\section{Introduction}
 The large CP violation in ${B\rightarrow J/\psi K_s}$
decay mode predicted by the standard model with 
Kobayashi-Masukawa (KM) scheme 
has been 
verified by ${B}$ factories at KEK (
High Energy Accelerator Research Organization) and 
Stanford Linear Accelerator Center (SLAC).
The standard model predicts the CP asymmetries
for ${B\rightarrow J/\psi K_s}$ and ${B\rightarrow \phi K_s}$
to be equal to ${\sin{2\phi_1}}$.
However, recent experimental data from Belle showed that
these asymmetries differ by nearly ${2 \sigma}$;
the averaged ${\sin{2\phi_1}}$ from Bell and BaBar
in ${B\to J/\psi K_s}$ system is 
${\sin{2\phi_1}=0.736 \pm 0.049}$ 
\cite{Eidelman:2004wy}
and in ${B\rightarrow \phi K_s}$ decay mode is
${\sin{2\phi_1}=0.06 \pm 0.33\pm 0.09}$ from Belle \cite{Abe:2004xp},
and
${\sin{2\phi_1}=0.50\pm 0.25^{+0.07}_{-0.04}}$ from BaBar
\cite{Aubert:2004dy}.
Experimental error is still large, so
the situation is inconclusive,
but if this result continues to hold,
it implies existence of new physics beyond the standard model.

In this paper, we want to concentrate on 
${B\rightarrow K^* \gamma }$ decay mode.
The decay mode has large a branching ratio,
so the experimental error
on the CP asymmetry has been getting small
and is now down to several percent.

\begin{equation*}
Br(B^0\rightarrow K^{*0} \gamma )=
\begin{cases}
(4.01\pm 0.21\pm 0.17)\times 10^{-5}&\mbox{\cite{Nakao:2004th}}\\
(3.92\pm 0.20\pm 0.24)\times 10^{-5}& \mbox{\cite{Aubert:2004te}}
\end{cases}
\end{equation*}
\begin{equation*}
Br(B^{\pm}\rightarrow K^{*\pm} \gamma )=
\begin{cases}
(4.25\pm 0.31\pm 0.24)\times 10^{-5}&\mbox{\cite{Nakao:2004th}}\\
(3.87\pm 0.28\pm 0.26)\times 10^{-5}& \mbox{\cite{Aubert:2004te}}
\end{cases}
\end{equation*}
\begin{equation*}
A_{CP}=
\begin{cases}
-0.015\pm 0.044\pm 0.012&\mbox{\cite{Nakao:2004th}}\\
-0.013\pm 0.036\pm 0.010& \mbox{\cite{Aubert:2004te}}
\end{cases}
\end{equation*}

CP Asymmetry is defined as 
\begin{eqnarray}
A_{CP}\equiv \frac{\Gamma(\bar{B}\to \bar{K^*}\gamma)-
\Gamma(B\to K^*\gamma)}{\Gamma(\bar{B}\to \bar{K^*}\gamma)+
\Gamma(B\to K^*\gamma)}~,
\end{eqnarray}
and in general, theoretical predictions of the CP asymmetries depend
less on hadronic parameters than those of branching ratios 
as many uncertainties cancel in the ratio.
So comparing predictions for CP asymmetries
within the standard model with experimental data
may be effective way to search for new physics.
Many authors have pointed out for some time, that
the CP asymmetry in this mode is very small.
We can easily understand why the asymmetry is so small.
In order to generate CP asymmetry, at least two amplitudes
with nonvanishing relative weak and strong phases
must interfere. 
This decay is mainly 
caused by an ${O_{7\gamma}}$ operator, and other contributions
which interfere with
this contribution are small and
the Cabibbo-Kobayashi-Masukawa quark-mixing matrix
(CKM) unitary triangle is crushed,
making the CP asymmetry in this decay mode very small.
However, if we were to look for new physics,
we need to be able to give a quantitative estimate of
the standard model contribution to the CP asymmetry.
For this purpose, we include small contributions 
which interfere with ${O_{7\gamma}}$,
including also the long distance contributions, for example, 
${B\rightarrow K^* J/\psi \rightarrow K^* \gamma}$.

Furthermore, the isospin breaking effect ${\Delta_{0+}}$ is also
very interesting because it's size and sign are sensitive
to the existence of physics beyond the standard model.
\begin{eqnarray}
\label{Isospin}
\Delta_{0+}
&\equiv & \frac{\Gamma(B^0\to K^{*0}\gamma)-\Gamma(B^+\to K^{*+}\gamma)}
{\Gamma(B^0\to K^{*0}\gamma)+\Gamma(B^+\to K^{*+}\gamma)}\nonumber\\
&=&
\frac{(\tau_{B^+}/\tau_{B^0})Br(B^0\to K^{*0}\gamma)-Br(B^+\to
K^{*+}\gamma)}
{(\tau_{B^+}/\tau_{B^0})Br(B^0\to K^{*0}\gamma)+Br(B^+\to
K^{*+}\gamma)}\nonumber\\
\end{eqnarray}
In order to test the standard model,
we need to know if the penguin contribution
within the standard model can explain the experimental
data.
Experiments show
${\Delta_{0+}=+0.012\pm 0.044\pm 0.026}$ in Bell \cite{Nakao:2004th}
and ${\Delta_{0-}=0.050\pm 0.045\pm 0.028\pm 0.024}$ 
in BaBar \cite{Aubert:2004te}.
More precise data will 
become available in the near future,
so the theoretical prediction of
it's size and sign of this asymmetry  
should be pinned down.

In this paper, we
calculate the branching ratio,
direct CP asymmetry, and isospin breaking effects
in ${B\to K^*\gamma}$ decay mode,
based on the standard model.
First, we briefly review the concept of
the pQCD in Sect.\ref{Outline of pQCD}, 
and in Sect.\ref{Effective Hamiltonian}, 
we show the effective Hamiltonian which 
causes   ${B\to K^*\gamma}$ decay.
Then 
we present the factorization formulas for
the ${B\rightarrow K^* \gamma}$ decay mode in Sect.\ref{Formulas},
and 
in Sect.\ref{Long distance}, we mention about the long distance
contributions.
Next  we will show the numerical results 
in Sect.\ref{Numerical},
and Sect.\ref{Conclusion} is our conclusion.
Finally in Appendix A, we present a brief review of pQCD.    

\section{Outline of pQCD}
\label{Outline of pQCD}
Theoretically, it is easy to analyze the inclusive ${B}$ meson decay
like ${B\rightarrow X_s \gamma}$
because we can estimate the decay width, for example,
by inserting the complete set for all possible
intermediate states.
The experimental and theoretical branching ratio of ${B\to X_s\gamma}$
are
\begin{eqnarray*}
Br(B\to X_s \gamma)^{\scriptsize\mbox{exp}}&=&(3.52^{+0.30}_{-0.28})\times 10^{-4},
\hspace{5mm}\mbox{\cite{Alexander:2005cx}}\\
Br(B\to X_s \gamma)^{\scriptsize\mbox{th}}&=&(3.57\pm 0.30)\times 10^{-4},
\hspace{5mm}\mbox{\cite{Buras:2002tp}}
\end{eqnarray*}
and this good agreement strongly constraints
new physics parameters.
However, 
inclusive decays are experimentally difficult to analyze
because all ${B\to X_s \gamma}$ candidates should be counted. 
If we can directly calculate the exclusive decay mode ${B\to K^* \gamma}$,
we ought to obtain many interesting results
to test the standard model or to search for new physics.

\begin{figure}[htbp]
\includegraphics[width=4.5cm]{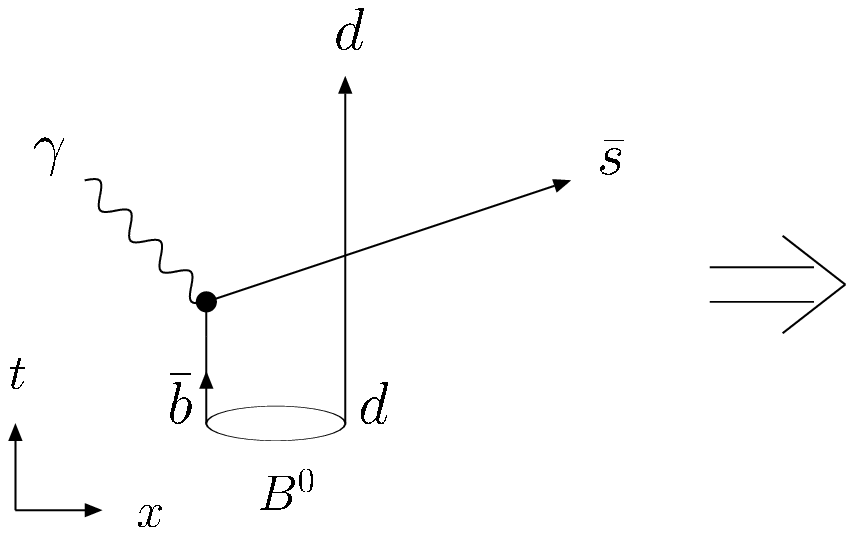}
\hspace{1mm}
\includegraphics[width=3.8cm]{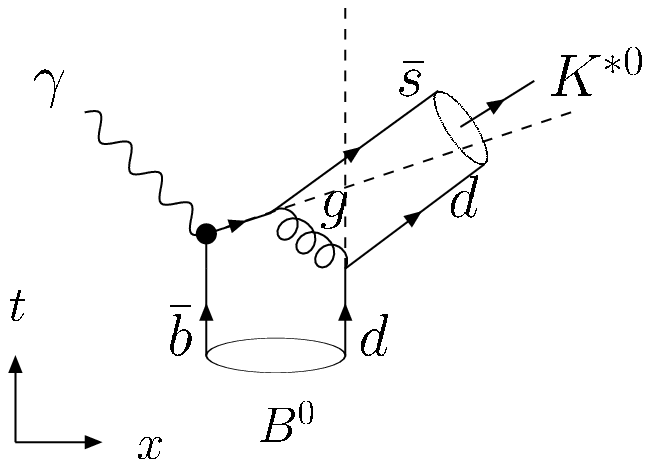}

\hspace{-0.5cm}
(a)\hspace{4.8cm}(b)
\caption{The left figure is no gluon exchange diagram.
 ${\bar s}$ and spectator  ${d}$ are not lines up 
to form an energetic ${K^*}$ meson.
In order to hadronize ${K^*}$ meson,
one gluon with large ${q^2}$ should be
exchanged.}
\label{factorization}
\end{figure}

Perturbative QCD
is one of the theoretical instrument for handling
the exclusive
decay modes.
The concept of pQCD is the factorization
between soft and hard dynamics.
In order to physically understand the pQCD approach,
we consider ${B^0}$ meson decays into ${K^{*0}}$ 
meson and ${\gamma}$ in the rest frame of the ${B^0}$ meson 
(Fig.\ref{factorization}).
The heavy ${\bar{b}}$ quark which has most of ${B^0}$ meson mass
is nearly static in this frame and
the other quark, which forms the ${B^0}$ meson together with the ${\bar{b}}$
quark, 
called the spectator quark, 
carries momentum of order ${O(\bar{\Lambda})=O(M_B-m_b)}$.
This ${\bar{b}}$ quark 
decays  into the light ${\bar{s}}$ quark and ${\gamma}$ 
through the electromagnetic penguin operator
and the decay products dash away 
back-to-back, 
with momentum of ${O(M_B/2)}$.
(This process is depicted in Fig.\ref{factorization}(a).)
${K^{*0}}$ meson is composed of ${\bar{s}}$ quark and a spectator quark.
In order for the fast moving ${\bar{s}}$ quark 
and slow moving spectator ${d}$  quark to form a
${K^{*0}}$ meson and nothing else,
the spectator quark
must be kicked by the gluon,
so that the ${\bar{s}}$ and ${d}$ quark have
more or less parallel momenta in the direction of ${K^{*0}}$.
(This process is depicted in Fig.\ref{factorization}(b).)
Since the invariant-mass square of this gluon is the order 
of ${O(\bar{\Lambda}M_B)}$, 
we can treat this decay process perturbatively.\\

\begin{figure}
\includegraphics[width=5cm]{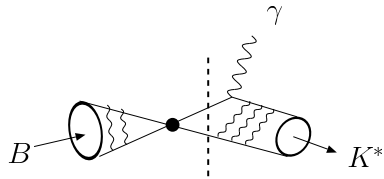}
\caption{Example of annihilation diagram which produces strong phase
through the branch-cut.}
\label{anni}
\end{figure}

There is also diagram shown in Fig.\ref{anni}.
This can also be computed in the pQCD approach.
The diagram
can be cut along the dotted line indicating the presence
of the physical intermediate state.
This results in a strong interaction phase
which can be computed.
The direct CP asymmetry is caused by interfering some
amplitudes which have relative weak and strong phase,
and it can be written in the form proportioning to
${\sin{(\theta_{w1} -\theta_{w2})}\sin{(\delta_{s1} -\delta_{s2})}}$:
in short, it depends on both weak and strong phases.
\begin{eqnarray*}
A(B\to f)&=&A_1 e^{i\theta_{w1}}e^{i\delta_{s1}}
+A_2 e^{i\theta_{w2}}e^{i\delta_{s2}}\\
A(\bar{B}\to \bar{f})&=&A_1 e^{-i\theta_{w1}}e^{i\delta_{s1}}+A_2
 e^{-i\theta_{w2}}e^{i\delta_{s2}}
\end{eqnarray*}

We can determine the strong phases by using the pQCD approach,
then we can extract the information about the weak phases
and examine the standard model.
A more detailed review for the pQCD approach is in Appendix \ref{pQCD}. 

\section{Kinematics for ${B\rightarrow K^* \gamma}$ decay mode}
\label{Effective Hamiltonian}
The effective Hamiltonian which
induces flavor-changing ${b\rightarrow s \gamma}$
transition is given by \cite{Buchalla:1995vs}
\begin{eqnarray}
H_{eff}&=&\frac{G_F}{\sqrt 2}
\Big[
\sum_{q=u,c}V_{qb}V_{qs}^*\left(C_{1}(\mu)O_1^{(q)}(\mu)+
C_{2}(\mu)O_2^{(q)}(\mu)\right)\nonumber\\
&&\hspace{1cm}-V_{tb}V_{ts}^*
\sum_{i=3\sim 8g}C_i(\mu)O_i(\mu)
\Big]+\mbox{h.c.}~,
\end{eqnarray}

\begin{eqnarray}
&&O_1^{(q)}=(\bar{s}_i q_j)_{V-A}(\bar{q}_j b_i)_{V-A},\nonumber\\
&&O_2^{(q)}=(\bar{s}_i q_i)_{V-A}(\bar{q}_j b_j)_{V-A},\nonumber\\
&&O_3=(\bar{s}_i b_i)_{V-A}\sum_q(\bar{q}_j q_j)_{V-A},\nonumber\\
&&O_4=(\bar{s}_i b_j)_{V-A}\sum_q(\bar{q}_j q_i)_{V-A},\\
&&O_5=(\bar{s}_i b_i)_{V-A}\sum_q(\bar{q}_j q_j)_{V+A},\nonumber\\
&&O_6=(\bar{s}_i b_j)_{V-A}\sum_q(\bar{q}_j q_i)_{V+A},\nonumber\\
&&O_{7\gamma}=\frac{e}{4{\pi}^2}\bar{s}_i\sigma^{\mu \nu}(m_sP_L
 +m_bP_R)b_iF_{\mu \nu},\nonumber\\
&&O_{8g}=\frac{g}{4{\pi}^2}\bar{s}_i\sigma^{\mu \nu}(m_sP_L
 +m_bP_R)T_{ij}^ab_jG_{\mu \nu}^a,\nonumber
\end{eqnarray}
where ${P^L_R=(1\mp\gamma^5)/2}$.
We define the ${B}$ meson and the ${K^{*}}$ meson 
momenta
${P_1}$ and ${P_2}$ in the light-cone coordinates
\begin{eqnarray}
p=(p^+,
 p^-,\vec{p}_{T})=\left(\frac{p^0+p^3}{\sqrt{2}},\frac{p^0-p^3}{\sqrt{2}},(p^1,p^2)\right)
\end{eqnarray}
within the ${B}$ meson rest frame as
\begin{eqnarray}
P_1 &=&(P_1^{+},P_1^{-},\vec P_{1T})=\frac{M_B}{\sqrt 2}(1,1,\vec 0_T),\label{1}\\
P_2 &=&(P_2^{+},P_2^{-},\vec P_{2T} )=
 \frac{M_B}{\sqrt{2}}(0,1,\vec 0_T),\label{2}
\end{eqnarray}
and photon and the ${K^*}$ meson transverse polarization vectors as
\begin{eqnarray}
\epsilon_{\gamma}^{\ast}(\pm)&=&\left(0,0,\frac{1}{\sqrt{2}}(\mp 1,-i)\right),\nonumber\\
\epsilon_{K^{\ast}}^{\ast}(\pm)&=&\left(0,0,\frac{1}{\sqrt{2}}(\pm 1,-i)\right).
\end{eqnarray}
Throughout this paper, we keep only terms of order
${r_{K^*}}$ in the computation of the numerator, where ${r_{K^*}=M_{K^*}/M_B}$.

The fractions of the momenta which have the spectator quarks 
in ${B}$ and ${K^{*}}$ meson are ${x_1=k_1^+/P_1^+}$ and
${x_2=k_2^-/P_2^-}$, so the momenta of these
spectator quarks 
are expressed as follows,
\begin{eqnarray}
k_1&=&(k_1^{+},k_1^{-},\vec k_{1T})=(\frac{M_B}{\sqrt 2}x_1,0,\vec{k_{1T}}),\label{3}\\
k_2&=&(k_2^{+},k_2^{-},\vec k_{2T} )=(0,\frac{M_B}{\sqrt 2}
x_2,\vec{k_{2T}}),\label{4}
\end{eqnarray}
then the ${b}$ and ${s}$ quark momenta are
${p_b=P_1-k_1}$ and ${p_s=P_2-k_2}$, and we neglect
the masses of the light quarks and identify the ${b}$ quark
mass with the ${B}$ meson mass in calculations of the hard 
scattering amplitudes. 
The term proportional to ${\Lambda_b=M_B-m_b}$
is generated by
higher order effects, so we included this effect
in our error estimate.

From here, we extract the formulas for decay amplitudes
caused by each operators,
\begin{eqnarray}
M&=&\langle F \mid H_{eff} \mid I \rangle \nonumber\\
&=&\frac{G_F}{\sqrt 2}\sum_i V_{CKM} C_i(\mu)\langle F \mid O_i(\mu) \mid I \rangle
\end{eqnarray}
and they
can be decomposed 
into scalar and pseudoscalar components as
\begin{eqnarray}
M=(\epsilon^{\ast}_{\gamma}\cdot \epsilon^{\ast}_{K^{\ast}})M^S +i
 \epsilon_{\mu \nu + -}\epsilon^{\ast \mu}_{\gamma}
\epsilon^{\ast \nu}_{K^{\ast}}M^P.
\end{eqnarray}

\section{Formulas}
\label{Formulas}
In this section, we want to show the explicit formulas
of the decay amplitudes caused by operators given in
Sect.\ref{Effective Hamiltonian}.
\subsection{${O_{7\gamma}}$ contribution}
\begin{figure}[htbp]
\includegraphics[width=3.6cm]{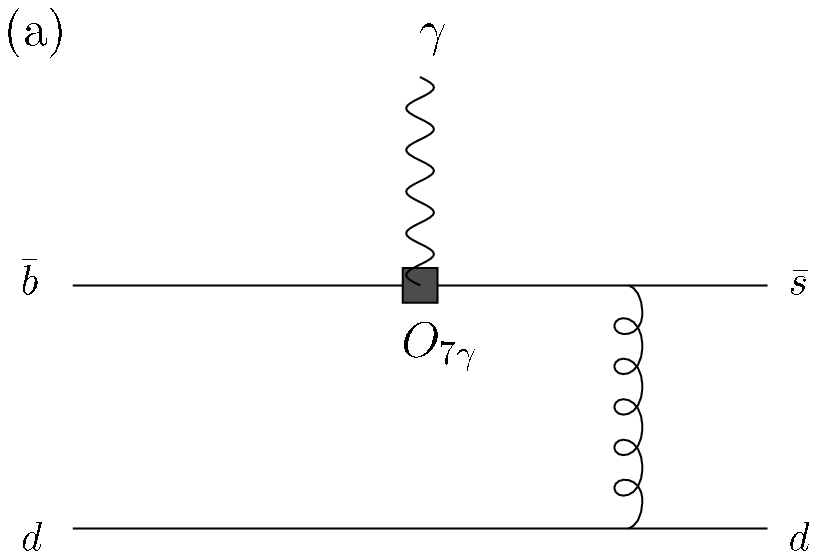}
\hspace{8mm}
\includegraphics[width=3.6cm]{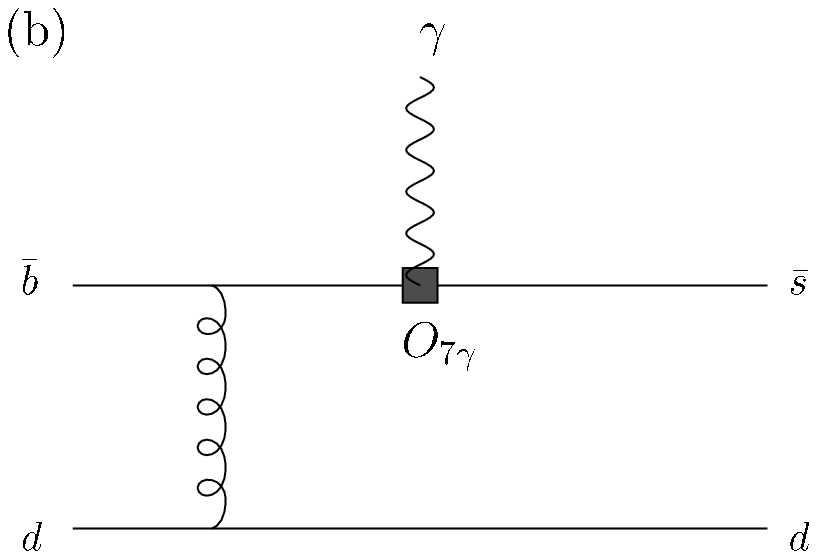}
\caption{Feynman diagrams of electromagnetic penguin operator
 ${O_{7\gamma}}$.
A photon is emitted through the  ${O_{7\gamma}}$
operator, and one hard gluon exchange is needed to form hadrons.}
\label{O7}
\end{figure}

If we define the common factor as
\begin{eqnarray}
F^{(0)}\equiv \frac{G_F}{\sqrt{2}}\frac{e}{\pi}V_{cb}^*V_{cs}C_F M_B^5 ,
\end{eqnarray}
where ${C_F}$ is color factor,
and ${\xi_i}$ as ${V_{ib}^*V_{is}/V_{cb}^*V_{cs}}$,
the decay amplitude
${M_{7\gamma}}$ in Fig.\ref{O7} can be
expressed as follows.

\begin{widetext}
\begin{eqnarray}
M_{7\gamma}^{S (a)}&=&-M_{7\gamma}^{P (a)}\nonumber\\
&=&-2\hspace{1mm}F^{(0)}\xi_t\int^{1}_{0}dx_1 dx_2 \int b_1 db_1 b_2 db_2
\phi_B(x_1,b_1)S_t(x_1)\alpha_s(t_7^a)
e^{\left[-S_B(t_7^a)-S_{K^*}(t_7^a)\right]}
C_7(t_7^a)\nonumber\\
& \times &H_7^{(a)}\left(A_7b_2,B_7b_1,B_7b_2\right)
r_{K^{\ast}}\left[\phi_{K^{\ast}}^v(x_2)
+\phi_{K^{\ast}}^a(x_2)\right]
\hspace{5mm}\left(t_7^a =\mbox{max}(A_7,B_7,1/b_1,1/b_2)\right)
\end{eqnarray}
\begin{eqnarray}
M_{7\gamma}^{S(b)}&=&-M_{7\gamma}^{P(b)}\nonumber\\
&=&-2\hspace{1mm}F^{(0)}\xi_t\int^{1}_{0}dx_1 dx_2 \int
 b_1db_1\hspace{1mm}
b_2 db_2
\phi_B(x_1,b_1)~S_t(x_2)\hspace{1mm}
\alpha_s(t_7^b)e^{\left[-S_B(t_7^b)-S_{K^*}(t_7^b)\right]}
C_7(t_7^b)\nonumber\\
&\times &H_7^{(b)}(A_7b_1,C_7b_1,C_7b_2)
\Big[(1-2x_2)r_{K^*}
 (\phi_{K^{\ast}}^v(x_2)
+\phi_{K^{\ast}}^a(x_2))
+(1+x_2)\phi_{K^{\ast}}^T(x_2)\Big]\nonumber\\
&&\hspace{25mm}\left(t_7^b =\mbox{max}(A_7,C_7,1/b_1,1/b_2)\right)
\end{eqnarray}
\begin{eqnarray}
&&\hspace{-1cm}H_7^{(a)}\left(A_7b_2,B_7b_1,B_7b_2\right)
\equiv 
K_0(A_7b_2)
\Big[\theta(b_1-b_2)K_0\left(B_7b_1\right)I_0\left(B_7b_2\right)
+\theta(b_2-b_1)K_0\left(B_7b_2\right)I_0\left(B_7b_1\right)\Big]
\end{eqnarray}
\begin{eqnarray}
H_7^{(b)}(A_7b_1,C_7b_1,C_7b_2) = H_7^{(a)}(A_7b_1,C_7b_1,C_7b_2)
\end{eqnarray}
\begin{eqnarray}
A_7^2= x_1x_2M_B^2,\hspace{5mm}B_7^2 = x_1{M_B}^2\hspace{5mm}C_7^2=  x_2 M_B^2 
\hspace{5mm}
\end{eqnarray}
\end{widetext}
Here ${K_0}$, ${I_0}$ are modified Bessel functions
which come from propagator integrations.
The meson wave functions are not calculable because of its 
nonperturbative feature. But they should be universal since 
they absorb long-distance
dynamics, so we can use the meson wave functions determined
by some approaches.
We use in this paper a model ${B}$ meson wave function
which is shown to give adequate form factors for ${B\to K\pi}$ decays
\cite{Bauer:fx, Keum:2000ph},
and ${K^*}$ meson determined by light-cone
QCD sum rule \cite{Ball:1998ff,Ball:2003sc}.
Their explicit formulas are shown in Appendix B.

\subsection{${O_{8g}}$ contribution}
Similarly, we can calculate  the ${O_{8g}}$ contributions as follows.
In these cases, a hard gluon is emitted through  the ${O_{8g}}$ operator
and glued to the spectator quark line (Fig.\ref{O8g}). 
In the following formulas, ${Q_q}$ express the electric
charge of the external quark: ${Q_u=2/3}$ and ${Q_d=Q_s=Q_b=-1/3}$.
\begin{widetext}
\begin{center}
\begin{figure}
\includegraphics[width=3.8cm]{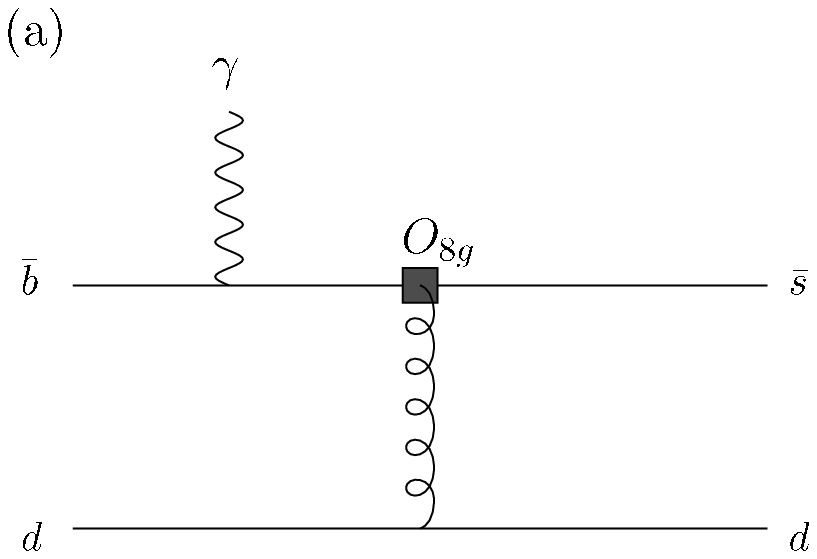}
\hspace{1cm}
\includegraphics[width=3.8cm]{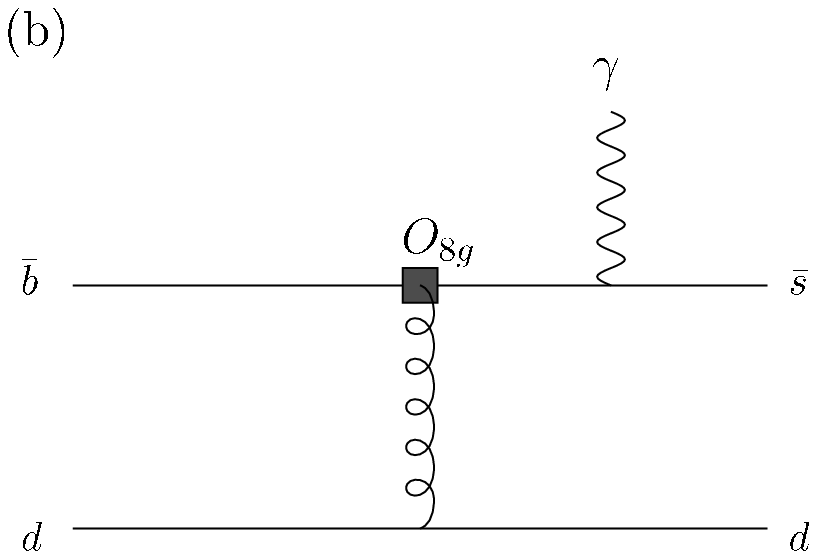}
\vspace{1cm}
\\
\includegraphics[width=3.8cm]{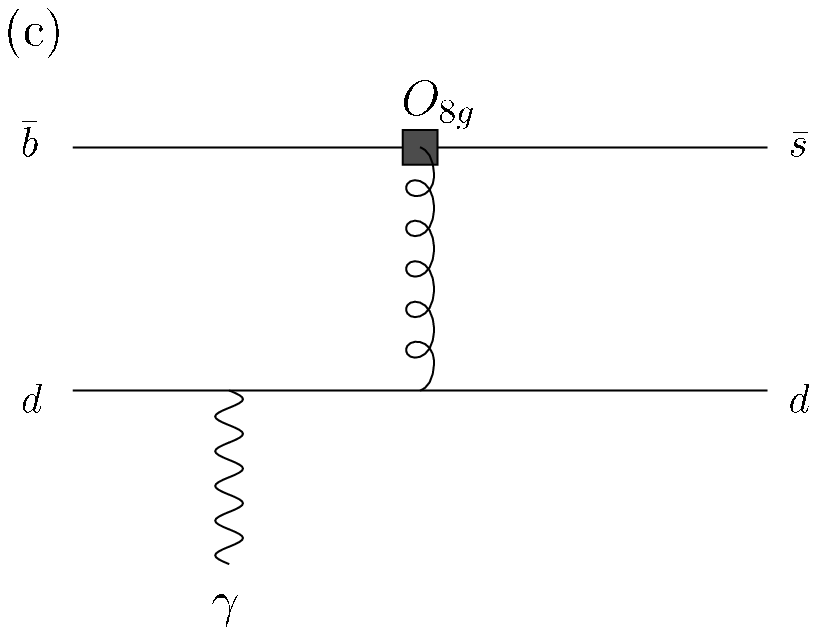}
\hspace{1cm}
\includegraphics[width=3.8cm]{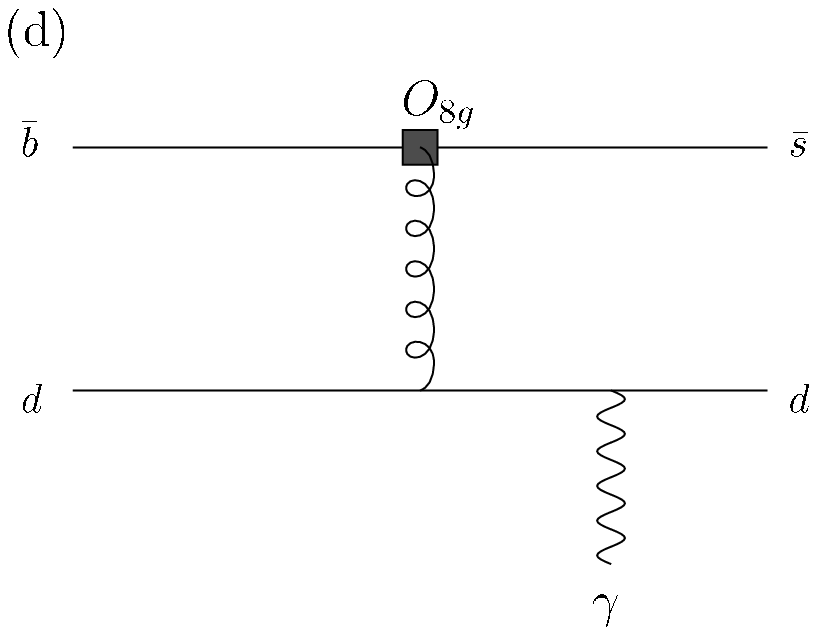}
\caption{Feynman diagrams of chromomagnetic penguin operator
 ${O_{8g}}$.
A hard gluon is emitted through  the ${O_{8g}}$ operator
and glued to the spectator quark line. 
Then photon is emitted
by bremsstrahlung of external quark lines. }
\label{O8g}
\end{figure}
\end{center}

\small{
\begin{eqnarray}
M_8^{S(a)}&=&-M_8^{P(a)}\nonumber\\
&=&-F^{(0)}\xi_t Q_b\int^1_0 dx_1 dx_2 \int b_1 db_1 
\hspace{1mm}b_2 db_2
\hspace{1mm}\phi_B(x_1,b_1)
S_t(x_1)\alpha_s(t_8^a)
e^{\left[-S_B(t_8^a)-S_{K^*}(t_8^a)\right]}C_8(t_8^a)H_8^{(a)}(A_8 b_2,B_8b_1,B_8b_2)\nonumber\\
&\times & [x_1 \phi_{K^*}^T(x_2)+r_{K^*}x_2(\phi_{K^*}^v(x_2)
+\phi_{K^*}^a(x_2))]
\hspace{5mm}\left(t_8^a
	       =\mbox{max}(A_8,B_8,1/b_1,1/b_2)\right)
\end{eqnarray}
\begin{eqnarray}
M_8^{S(b)}&=&-M_8^{P(b)}\nonumber\\
&=&-F^{(0)}\xi_t Q_s
\int^1_0 dx_1 dx_2 \int b_1 db_1 \hspace{1mm}b_2 db_2
\hspace{1mm}\phi_B(x_1,b_1)
S_t(x_2)\alpha_s(t_8^b)\hspace{1mm}
e^{\left[-S_B(t_8^b)-S_{K^*}(t_8^b)\right]}C_8(t_8^b)H_8^{(b)}(A_8 b_1,C_8b_1,C_8b_2)
\nonumber\\
&\times & 
[-3x_2r_{K^*}(\phi_{K^*}^v(x_2)+\phi_{K^*}^a(x_2))
+(2x_2-x_1)\phi_{K^*}^T(x_2)]
\hspace{5mm}\left(t_8^b=\mbox{max}(A_8,C_8,1/b_1,1/b_2)\right)
\end{eqnarray}
\begin{eqnarray}
M_8^{S(c)}&=&-M_8^{P(c)}\nonumber\\
&=&-F^{(0)}\xi_t Q_q
\int^1_0 dx_1 dx_2 \int b_1 db_1 \hspace{1mm}b_2 db_2
\hspace{1mm}
\phi_B(x_1,b_1)
S_t(x_1)\alpha_s(t_8^c)\hspace{1mm}
e^{\left[-S_B(t_8^c)-S_{K^*}(t_8^c)\right]}
C_8(t_8^c)H_8^{(c)}(\sqrt{|A_8'^2|}b_2,D_8b_1,D_8b_2)
\nonumber\\
&&\hspace{-5mm}\times 
[-x_1\phi_{K^*}^T(x_2)+x_2r_{K^*}(\phi_{K^*}^v(x_2)
+\phi_{K^{\ast}}^a(x_2))]
\hspace{5mm}\left(t_8^c= \mbox{max}(\sqrt{\mid A_8'^2\mid}
	       ,D_8,1/b_1,1/b_2)\right)
\end{eqnarray}
\begin{eqnarray}
M_8^{S(d)}
&=&-F^{(0)}\xi_t Q_q
\int^1_0 dx_1 dx_2 \int b_1 db_1 \hspace{1mm}b_2 db_2
\hspace{1mm}\phi_B(x_1,b_1)
S_t(x_2)\alpha_s(t_8^d)\hspace{1mm}
e^{\left[-S_B(t_8^d)-S_{K^{\ast}}(t_8^d)\right]}C_8(t_8^d)H_8^{(d)}
(\sqrt{|A_8'^2|}b_1,E_8b_1,E_8b_2)
\nonumber\\
&\times & 
[~ 6x_2
r_{K^{\ast}}\phi_{K^{\ast}}^v(x_2)
+(2+x_2-x_1)\phi_{K^{\ast}}^T(x_2)]
\end{eqnarray}
\begin{eqnarray}
M_8^{P(d)}
&=& F^{(0)}\xi_t Q_q
\int^1_0 dx_1 dx_2 \int b_1 db_1 \hspace{1mm}b_2 db_2
\hspace{1mm}\phi_B(x_1,b_1)
S_t(x_2)\alpha_s(t_8^d)
\hspace{1mm}e^{\left[-S_B(t_8^d)-S_{K^{\ast}}(t_8^d)\right]}
C_8(t_8^d)H_8^{(d)}
(\sqrt{|A_8'^2|}b_1,E_8b_1,E_8b_2)
\nonumber\\
&\times & 
[~(2+x_2-x_1)\phi_{K^{\ast}}^T(x_2)+
6x_2r_{K^{\ast}}\phi_{K^{\ast}}^a(x_2)]
\hspace{5mm}\left(t_8^d= \mbox{max}(\sqrt{|A_8'^2|} ,E_8,1/b_1,1/b_2)\right)
\end{eqnarray}}
\begin{eqnarray}
&&H_8^{(a)}(A_8 b_2,B_8b_1,B_8b_2)
\equiv K_0(A_8b_2) 
\Big[
\theta(b_1-b_2)K_0(B_8b_1)I_0(B_8b_2)
+(b_1\leftrightarrow b_2)
\Big]\\
&&H_8^{(b)}(A_8 b_1,C_8b_1,C_8b_2)
\equiv \frac{i\pi}{2}K_0(A_8b_1)
\Big[
\theta(b_1-b_2)H_0^{(1)}(C_8b_1)J_0(C_8b_2)
+(b_1 \leftrightarrow b_2)
\Big]\\
&&H_8^{(c)}(\sqrt{|A_8'^2|}b_2,D_8 b_1,D_8 b_2)
\equiv \theta(A_8'^2)\hspace{1mm}
K_0(\sqrt{|A_8'^2|}b_2)
\Big[
\theta(b_1-b_2)K_0(D_8 b_1)I_0(D_8 b_2)
+(b_1\leftrightarrow
b_2)\Big]\nonumber\\
&&\hspace{4cm}+\theta(-A_8'^2)\hspace{1mm}i\frac{\pi}{2}
H_0^{(1)}(\sqrt{|A_8'^2|} b_2) 
\Big[
\theta(b_1-b_2)K_0(D_8 b_1)I_0(D_8 b_2)+(b_1\leftrightarrow
b_2)
\Big]\\
&&H_8^{(d)}(\sqrt{|A_8'^2|}b_1,E_8b_1,E_8b_2)
\equiv \theta(A_8'^2)\hspace{1mm}
i\frac{\pi}{2}
K_0(\sqrt{|A_8'^2|}b_1)
\left[\theta(b_1-b_2)H_0^{(1)}(E_8b_1)J_0(E_8 b_2)
+(b_1 \leftrightarrow b_2)\right]\nonumber\\
&&\hspace{4cm}-\theta(-A_8'^2)\hspace{1mm}\left(\frac{\pi}{2}\right)^2
H_0^{(1)}(\sqrt{|A_8'^2}|b_1)
\left[\theta(b_1-b_2)H_0^{(1)}(E_8b_1)J_0(E_8 b_2)
+(b_1 \leftrightarrow b_2)\right] 
\end{eqnarray}
\begin{eqnarray}
A_8^2 = x_1 x_2 M_B^2,\hspace{3mm}B_8^2=M_B^2(1+x_1),\hspace{3mm}
C_8^2=M_B^2(1-x_2),\hspace{3mm}
A_8'^2=(x_1-x_2)M_B^2,\hspace{3mm}
D_8^2 =x_1{M_B}^2,\hspace{3mm}E_8^2 =x_2 {M_B}^2\end{eqnarray}

\subsection{Loop contributions}
\subsubsection{Quark line photon emission}
Next we want to mention about charm and up quark
penguin contributions (Fig.\ref{cloop}).
The subtitle like ``Quark line photon emission'' means
that a photon is emitted through the external quark lines.
\begin{figure}
\includegraphics[width=3.8cm]{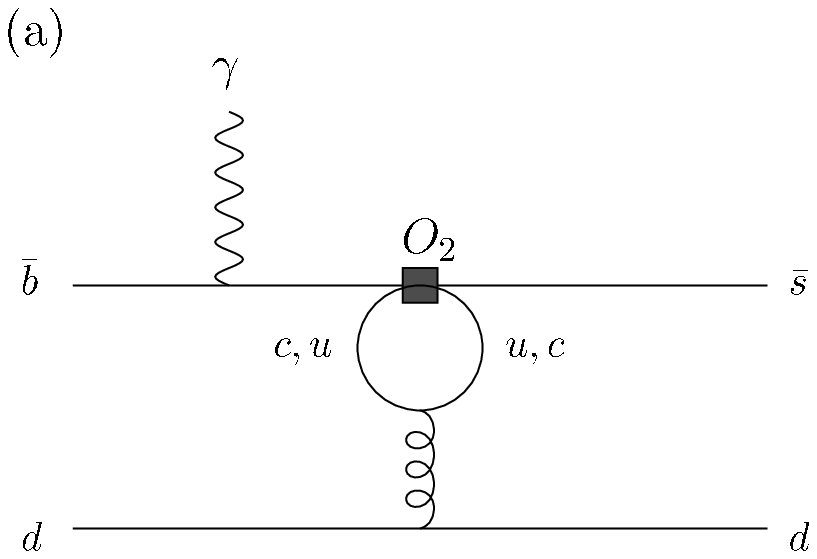}
\hspace{6mm}
\includegraphics[width=3.8cm]{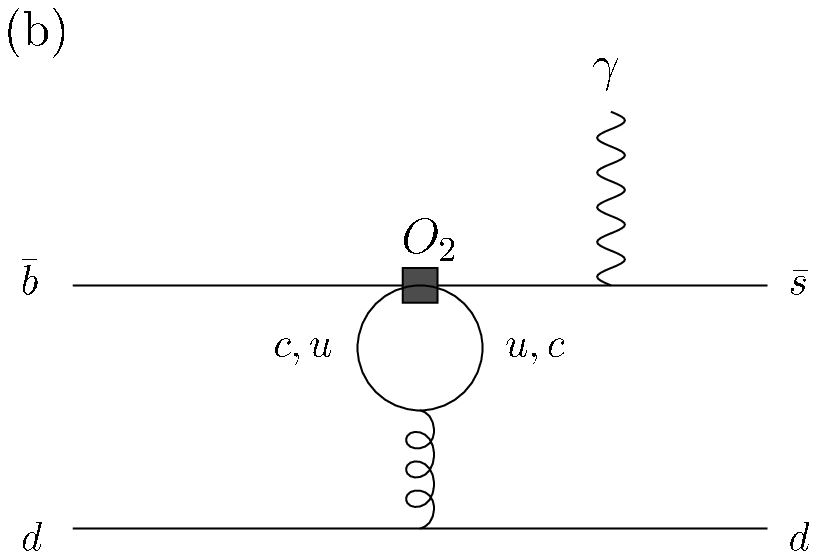}
\vspace{1cm}
\\
\includegraphics[width=3.8cm]{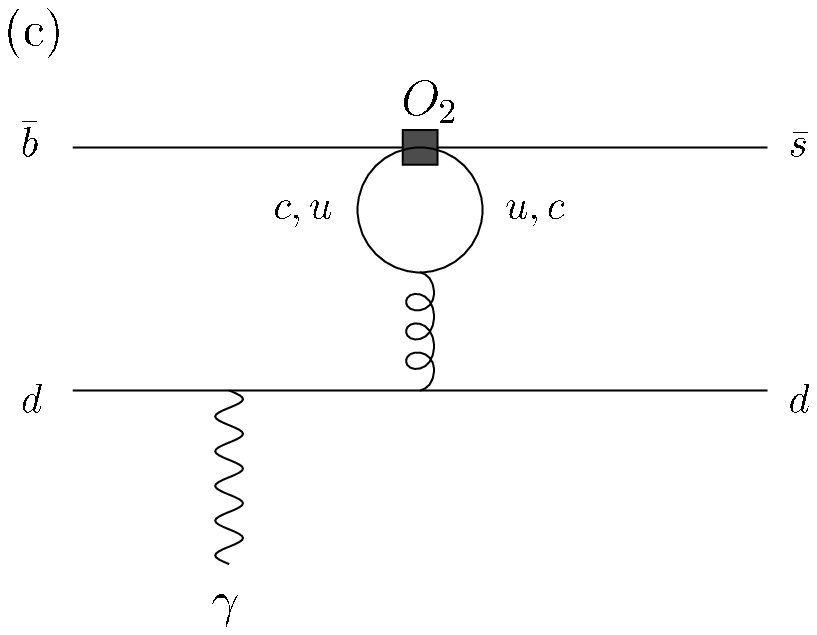}
\hspace{6mm}
\includegraphics[width=3.8cm]{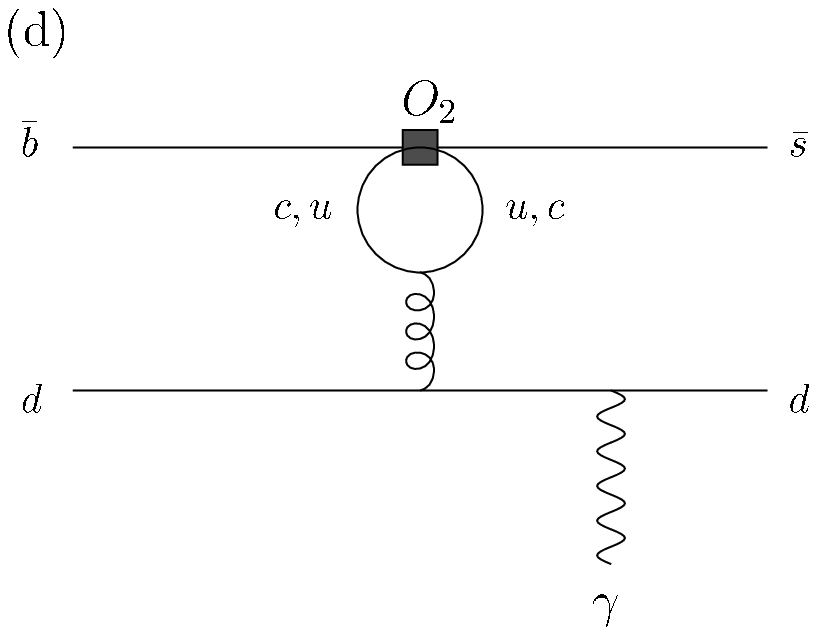}
\caption{Feynman diagrams of ``Quark line photon emission''.
The charm or up loop go to gluon and attach to the spectator quark
line.
A photon is emitted through the external quark lines.}
\label{cloop}
\end{figure}
We define the ${c}$ and ${u}$ loop function in order that
the ${b\to s g}$ vertex can be expressed as
${\bar{s}\gamma^{\mu}(1-\gamma^5)I_{\mu\nu}b}$.
It has the gauge invariant form \cite{Bander:px} and the explicit
formula is as follows,
\begin{eqnarray}
I_{\mu\nu}^a&=&\frac{gT^a}{2\pi^2}(k^2 g_{\mu\nu}-k_{\mu}k_{\nu})
\int^1_0 dx
x(1-x)
\left[1+\log\left[\frac{m_i^2-x(1-x)k^2}{t^2}\right]\right]\nonumber\\
&=&-\frac{gT^a}{8\pi^2}(k^2 g_{\mu\nu}-k_{\mu}k_{\nu})\left[G(m_i^2,k^2,t)-\frac{2}{3}\right],
\end{eqnarray}
where ${k}$ is the gluon momentum and ${m_i}$ is the loop internal quark
mass.
\begin{eqnarray}
G(m_i^2,k^2,t)
&=&
\theta(-k^2)\frac{2}{3}\Big[
\frac{5}{3}+\frac{4m_i^2}{k^2}-\ln{\frac{m_i^2}{t^2}}
+\left(1+\frac{2m_i^2}{k^2}\right)
\sqrt{1-\frac{4m_i^2}{k^2}}
\ln{\frac{\sqrt{1-4m_i^2/k^2}-1}{\sqrt{1-4m_i^2/k^2}+1}}
\Big]\nonumber\\
&&+\theta(k^2)\theta(4m_i^2-k^2)
\frac{2}{3}\Big[
\frac{5}{3}+\frac{4m_i^2}{k^2}-\ln{\frac{m_i^2}{t^2}}
-2\left(1+\frac{2m_i^2}{k^2}\right)
\sqrt{\frac{4m_i^2}{k^2}-1}\arctan{\left(\frac{1}{\sqrt{4m_i^2/k^2-1}}\right)}
\Big]\nonumber\\
&&+\theta(k^2-4m_i^2)
\frac{2}{3}\Big[
\frac{5}{3}+\frac{4m_i^2}{k^2}-\ln{\frac{m_i^2}{t^2}}
+\left(1+\frac{2m_i^2}{k^2}\right)
\sqrt{1-\frac{4m_i^2}{k^2}}
\left[
\ln{\frac{1-\sqrt{1-4m_i^2/k^2}}{1+\sqrt{1-4m_i^2/k^2}}}
+i\pi
\right]
\Big]\nonumber\\
\end{eqnarray}
The loop function ${G}$ has the dependence of gluon momentum square of
${k^2}$. But there is no singularity when we take the
limit of ${k\rightarrow 0}$, so we can neglect ${k_T}$
components of ${k^2}$ in the loop function ${G}$.

Then the ``Quark line photon emission'' contributions can be expressed
as follows.
\small{
\begin{eqnarray}
M_{1i}^{S(a)}&=&M_{1i}^{P(a)}\nonumber\\
&=&\frac{Q_b}{2}F^{(0)}\xi_{i}
\int^1_0 dx_1 dx_2 \int b_1 db_1 b_2 db_2 \phi_B(x_1,b_1)
 C_2(t_2^a)\alpha_s(t_2^a)
S_t(x_1)e^{\left[-S_B(t_2^a)-S_{K^*}(t_2^a)\right]}
H_2^{(a)}(A_2b_2,B_2b_1,B_2b_2)\nonumber\\
&&\times \Big[G(m_i^2,-A_2^2,t_2^a)-\frac{2}{3}\Big]
x_1x_2r_{K^*}(\phi_{K^*}^v(x_2)-\phi_{K^*}^a(x_2))
\hspace{5mm}\left(t_2^a=\mbox{max}(A_2,B_2,\hspace{1mm}1/b_1,1/b_2)\right)
\end{eqnarray}
\begin{eqnarray}
M_{1i}^{S(b)}
&=&-M_{1i}^{P(b)}\nonumber\\
&=&-\frac{Q_s}{2}F^{(0)}\hspace{1mm}\xi_{i}
\int^1_0 dx_1 dx_2 \int b_1 db_1 b_2 db_2 \phi_B(x_1,b_1)
C_2(t_2^b)\alpha_s(t_2^b)S_t(x_2)e^{\left[-S_B(t_2^b)-S_{K^*}(t_2^b)\right]}
H_2^{(b)}(A_2b_1,C_2b_1,C_2b_2^b)\nonumber\\
&&\times \Big[G(m_i^2,-A_2^2,t_2^b)-\frac{2}{3}\Big]
\Big[
x_2^2r_{K^*}(\phi_{K^*}^{v}(x_2)+\phi_{K^*}^{a}(x_2))
+3x_1x_2\phi_{K^*}^T(x_2)
\Big]
\hspace{5mm}\left(t_2^b=\mbox{max}(A_2,C_2,\hspace{1mm}1/b_1,1/b_2)\right)
\end{eqnarray}
\begin{eqnarray}
M_{1i}^{S(c)}&=&-M_{1i}^{P(c)}\nonumber\\
&=&\frac{Q_q}{2}F^{(0)}\hspace{1mm}\xi_{i}
\int^1_0 dx_1 dx_2 \int b_1 db_1 b_2 db_2 \phi_B(x_1,b_1)
C_2(t_2^c)\alpha_s(t_2^c)S_t(x_1)
e^{\left[-S_B(t_2^c)-S_{K^*}(t_2^c)\right]}
H_2^{(c)}(\sqrt{|A_2^{'2}|}b_2,D_2b_1,D_2b_2)\nonumber\\ 
&&\times \Big[G(m_i^2,-A_2^{'2},t_2^c)
-\frac{2}{3}\Big]
\Big[
x_2r_{K^*}(\phi_{K^*}^v(x_2)+\phi_{K^*}^a(x_2))
-x_1\phi_{K^*}^T(x_2)
\Big]
\hspace{5mm}\left(t_2^c=\mbox{max}(\sqrt{|A_2^{'2}|},D_2
 ,\hspace{1mm}1/b_1,1/b_2)\right)
\end{eqnarray}
\begin{eqnarray}
M_{1i}^{S(d)}&=&
\frac{Q_q}{2}F^{(0)}\hspace{1mm}\xi_{i}
\int^1_0 dx_1 dx_2 \int b_1 db_1 b_2 db_2 \phi_B(x_1,b_1)
C_2(t_2^d)
\alpha_s(t_2^d)S_t(x_2)e^{\left[-S_B(t_2^d)-S_{K^*}(t_2^d)\right]}
H_2^{(d)}(\sqrt{|A_2^{'2}|}b_1,E_2b_1,E_2b_2)\nonumber\\
&&\times \Big[G(m_i^2,-A_2^{'2},t_2^d)
-\frac{2}{3}\Big]
\Big[x_2r_{K^*}[3(1+x_2)\phi_{K^*}^v(x_2)
-(1-x_2)\phi_{K^*}^a(x_2)]
+3(x_2-x_1)\phi_{K^*}^T(x_2)
\Big]
\end{eqnarray}
\begin{eqnarray}
M_{1i}^{P(d)}&=&
-\frac{Q_q}{2}F^{(0)}\hspace{1mm}\xi_{i}
\int^1_0 dx_1 dx_2 \int b_1 db_1 b_2 db_2 \phi_B(x_1,b_1)
C_2(t_2^d)
\alpha_s(t_2^d)S_t(x_2)e^{\left[-S_B(t_2^d)-S_{K^*}(t_2^d)\right]}
H_2^{(d)}(\sqrt{|A_2^{'2}|}b_1,E_2b_1,E_2b_2)\nonumber\\
&&\times \Big[G(m_i^2,-A_2^{'2},t_2^d)-\frac{2}{3}\Big]
\Big[
x_2r_{K^*}[-(1-x_2)\phi_{K^*}^v(x_2)
+3(1+x_2)\phi_{K^*}^a(x_2)]
+3(x_2-x_1)\phi_{K^*}^T(x_2)
\Big]\nonumber\\
&&\hspace{4cm}
\left(t_2^d=\mbox{max}(\sqrt{|A_2^{'2}|},E_2,\hspace{1mm}1/b_1, 1/b_2)\right)
\end{eqnarray}}
\begin{eqnarray}
H_2^{(a)}(A_2 b_2,B_2b_1,B_2b_2)&=& H_8^{(a)}(A_8b_2, B_8b_1, B_8b_2) 
\nonumber\\
H_2^{(b)}(A_2b_1,C_2b_1,C_2b_2)&=&
H_8^{(b)}(A_8b_1, C_8b_1, C_8b_2)
\nonumber\\
H_2^{(c)}(\sqrt{|A_2^{'2}|}b_2,D_2b_1,D_2b_2)&=&
H_8^{(c)}(\sqrt{|A_8^{'2}|}b_2, D_8b_1,D_8b_2)
\\
H_2^{(d)}(\sqrt{|A_2^{'2}|}b_1,E_2b_1,E_2b_2)
&=& 
H_8^{(d)}(\sqrt{|A_8^{'2}|}b_1, E_8b_1, E_8b_2)
\nonumber
\end{eqnarray}
\begin{eqnarray}
A_2^2=x_1x_2M_B^2,\hspace{3mm}B_2^2=(1+x_1)M_B^2,\hspace{3mm}C_2^2=(1-x_2)M_B^2,\hspace{3mm}
A_2^{'2}=(x_1-x_2)M_B^2,\hspace{3mm}D_2^2=x_1 M_B^2,
\hspace{3mm}E_2^2=x_2M_B^2
\end{eqnarray}

\subsubsection{Loop line photon emission}
Next we consider the ``Loop line photon emission'':
a photon is emitted through the ${c}$ or ${u}$ 
quark loop line.
\begin{figure}
\includegraphics[width=4.5cm]{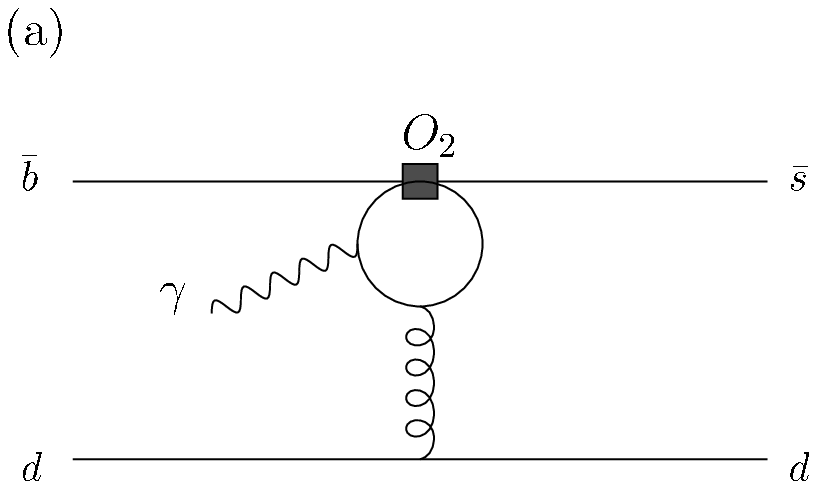}
\hspace{1cm}
\includegraphics[width=4.5cm]{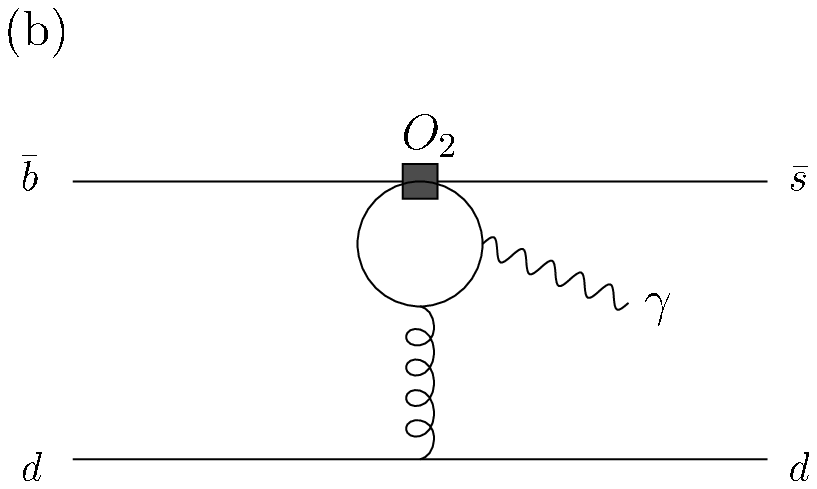}
\caption{The Feynman diagrams of ``Loop line photon emission''.
The charm or up loop go to gluon and attach to the spectator quark line.
A photon is emitted by bremsstrahlung of loop quark line.}
\label{loop-photon}
\end{figure}
We sum up Fig.\ref{loop-photon}(a) and \ref{loop-photon}(b),
the ${b\to sg\gamma}$ decay amplitude
is expressed as 
\begin{eqnarray}
A(b\to sg\gamma)=\epsilon_{\gamma}^{\mu}(q)
\epsilon_a^{\nu}(k)
\hspace{1mm}\bar{s}(p')I_{\mu\nu}^ab(p),
\end{eqnarray}
where vertex function ${I_{\mu\nu}}$ is defined as follows \cite{Liu:1990yb,Simma:nr,Li:1998tk},
\begin{eqnarray}
I_{\mu\nu}^a&=&F_1^a\Big[
(k\cdot q)\epsilon_{\mu\nu\rho\sigma}\hspace{1mm}
(q^{\rho}-k^{\rho})\gamma^{\sigma}
+q_{\nu}\epsilon_{\mu\rho\sigma\delta}\hspace{1mm}
q^{\rho}k^{\sigma}\gamma^{\delta}
-k_{\mu}\epsilon_{\nu\rho\sigma\delta}\hspace{1mm}
q^{\rho}k^{\sigma}\gamma^{\delta}
\Big]L\nonumber\\
&&
+F_2^a\Big[
k_{\nu}\epsilon_{\mu\rho\sigma\delta}\hspace{1mm}q^{\rho}k^{\sigma}
\gamma^{\delta}
+k^2\hspace{1mm}\epsilon_{\mu\nu\rho\sigma}\hspace{1mm}
 q^{\rho}\gamma^{\sigma}
\Big]L,
\end{eqnarray}
\begin{eqnarray}
F_1^a&=&-\frac{i2\sqrt{2}}{3\pi^2}egT^aG_F\int^1_0 dx 
\int^{1-x}_0 dy
\frac{xy}{m_i^2-2xy(k\cdot q)-k^2x(1-x)},\\
F_2^a &=&-\frac{i2\sqrt{2}}{3\pi^2}egT^aG_F\int^1_0 dx
\int^{1-x}_0 dy
\frac{x(1-x)}{m_i^2-2xy(k\cdot q)-k^2x(1-x)},
\end{eqnarray}
where ${L=(1-\gamma^5)/2}$, ${k}$ is the gluon momentum and
${q}$ is the photon one.
Then the amplitudes can be expressed as follows:

\begin{eqnarray}
M_{2i}^{S}
&=&-\frac{4}{3}F^{(0)}\xi_i
\int^1_0 dx \int^{1-x}_0 dy
\int^1_0 dx_1 dx_2 \int b_1 db_1 
\phi_B(x_1,b_1)C_2(t_2)\alpha_s(t_2)
e^{\left[-S_B(t_2)-S_{K^*}(t_2)\right]}
H_2(b_1A,b_1\sqrt{|B^2|})\nonumber\\
&&\times \frac{1}{xyx_2 M_B^2-m_i^2}
\Big[xyx_2\Big[(1-2x_2)r_{K^*}\phi_{K^*}^v(x_2)
-(1+2x_1)
\phi_{K^*}^T(x_2)
+r_{K^*}\phi_{K^*}^a(x_2)
\Big]\nonumber\\
&&\hspace{4cm}+x(1-x)
\Big[
x_2^2r_{K^*}(\phi_{K^*}^v(x_2)
+\phi_{K^*}^a(x_2))+3x_1x_2\phi_{K^*}^T(x_2)
\Big]\Big]
\nonumber\\
\end{eqnarray}
\begin{eqnarray}
M_{2i}^{P}
&=&\frac{4}{3}F^{(0)}\xi_i
\int^1_0 dx \int^{1-x}_0 dy
\int^1_0 dx_1 dx_2 \int b_1 db_1
\phi_B(x_1,b_1)C_2(t_2)\alpha_s(t_2)
e^{\left[-S_B(t_2)-S_{K^*}(t_2)\right]}
H_2(b_1A,b_1\sqrt{|B^2|})\nonumber\\
&&\times \frac{1}{xyx_2 M_B^2-m_i^2}
\Big[xyx_2\Big[
r_{K^*}\phi_{K^*}^v(x_2)
-(1+2x_1)\phi_{K^*}^T(x_2)
+(1-2x_2)r_{K^*}\phi_{K^*}^a(x_2)
\Big]\nonumber\\
&&\hspace{4cm}+x(1-x)
[x_2^2
r_{K^*}(\phi_{K^*}^v(x_2)+\phi_{K^*}^a(x_2))+3x_1x_2\phi_{K^*}^T(x_2)]
\Big]\nonumber\\
&&
\hspace{3cm}\left(t_2=\mbox{max}(A,\sqrt{|
B^2|},\hspace{1mm}1/b_1)\right)
\end{eqnarray}
\begin{eqnarray}
H_2(b_1A,b_1\sqrt{|B^2|})
&\equiv &
K_0(b_1A)-K_0(b_1 \sqrt{|B^2|})\hspace{5mm}(B^2 \geq 0)\nonumber\\
&\equiv &K_0(b_1A)-i\frac{\pi}{2}H_0(b_1 \sqrt{|B^2|})\hspace{3mm}(B^2 < 0)
\end{eqnarray}
\begin{eqnarray}
A^2=x_1x_2M_B^2,\hspace{5mm}
B^2=x_1x_2M_B^2 -\frac{y}{1-x}x_2M_B^2+\frac{m_i^2}{x(1-x)}
\end{eqnarray}

In general, it is hard to estimate the ${u}$ loop contributions
accurately because of the nonperturbative hadronic uncertainties.
For ${k^2<1}$GeV, nonperturbative correction to the
${u}$ quark loop shown in Fig.\ref{coupling}
is large and in fact ${u\bar{u}}$ pair may be
better represented by resonances. On the other hand
if ${k^2}$ is large, the perturbative computation is expected to be
reliable.

\begin{figure}
\includegraphics[width=4cm]{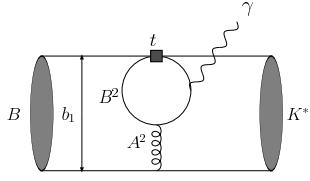}
\caption{The energy scale of the loop contribution is
 ${t=\mbox{max}(A,\sqrt{|B|^2},1/b_1)}$, where ${A^2}$ is gluon momentum
 , ${B^2}$ is the loop momentum, and ${b_1}$
is the transverse interval between the quark and
antiquark pair in the ${B}$ meson.}
\label{coupling}
\end{figure}

In the pQCD approach, the factorization energy scale ${t}$
is determined at each point of the integration,
i.e. for each point ${(x_1,x_2,b_1,b_2)}$.
Then these variables are integrated over
the entire physical region.
So for each point ${(x_1,x_2,b_1,b_2)}$,
 ${\alpha_s(t)}$
can be determined.
Thus we can observe the contribution to the
amplitude as a function of ${\alpha_s(t)}$.
Figures \ref{c} and \ref{u} 
show the distribution of ${\alpha_s(t)}$
for a diagram with the ${c}$ quark loop,
and ${u}$ quark loop, respectively.
We can see that 
the major part of the
${c}$ loop contribution
comes from 
a perturbative region, 
on the other hand ${u}$ loop contribution
includes also a nonperturbative region.
Since ${M_{2u}^S}$ and ${M_{2u}^P}$ gets considerable
contributions from the nonperturbative region,
we introduce 100\% theoretical error
for these amplitudes.

\begin{figure}[htbp]
\includegraphics[width=5.8cm]{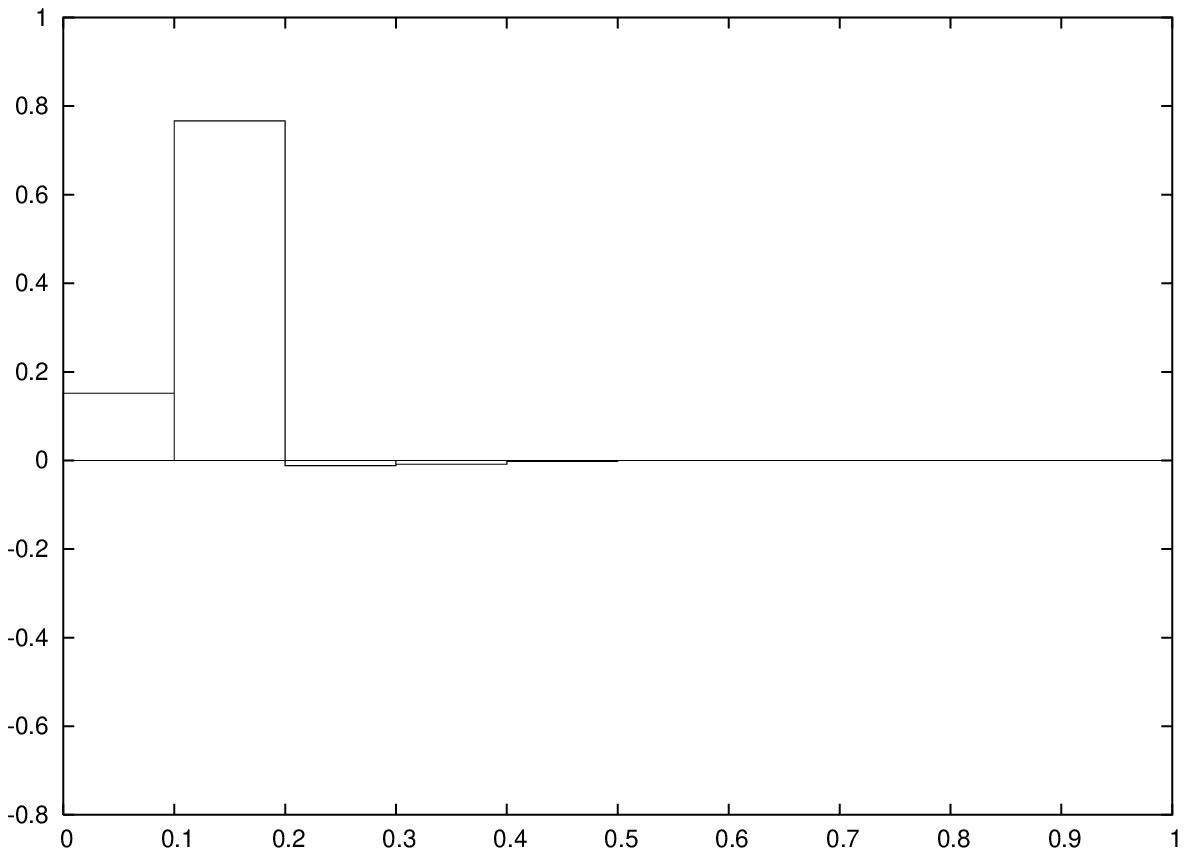}
\hspace{3mm}
\includegraphics[width=5.8cm]{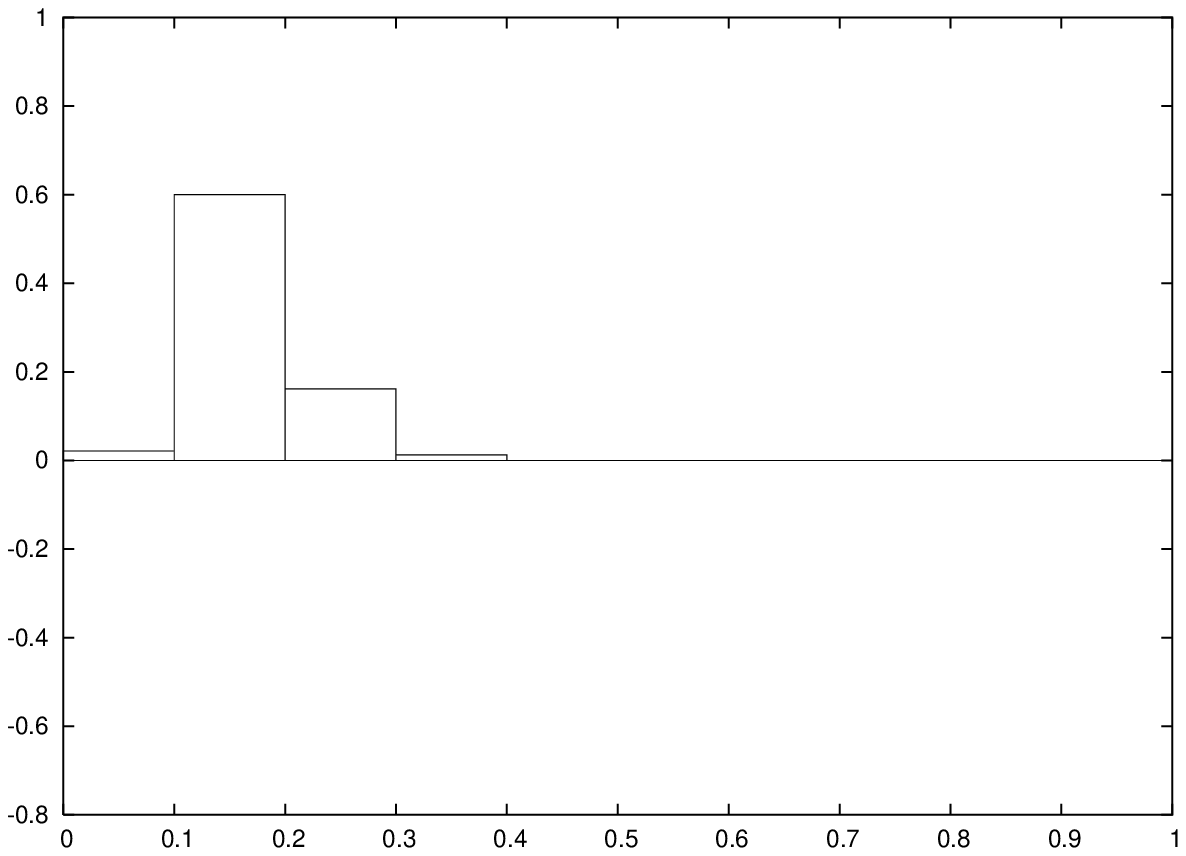}
\caption{The horizontal line is ${\alpha_s(t)/\pi}$ and the vertical
axis is the contribution from the each energy region 
in ${\alpha_s}$ to the total decay
 amplitude. The left figure is the real part and right one is
imaginary part
of the ${c}$ loop contribution.
These figures show that we can compute
the ${c}$  quark loop contribution perturbatively. }
\label{c}

\includegraphics[width=5.8cm]{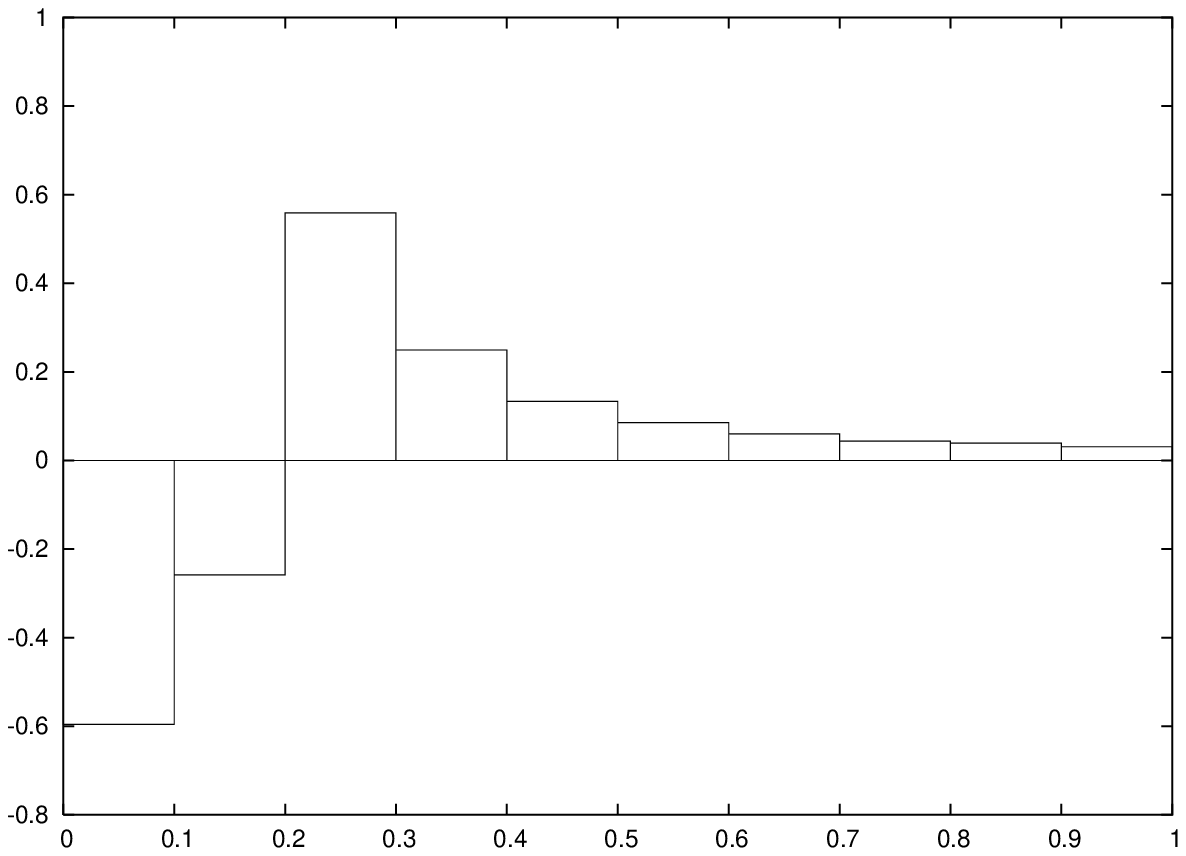}
\hspace{3mm}
\includegraphics[width=5.8cm]{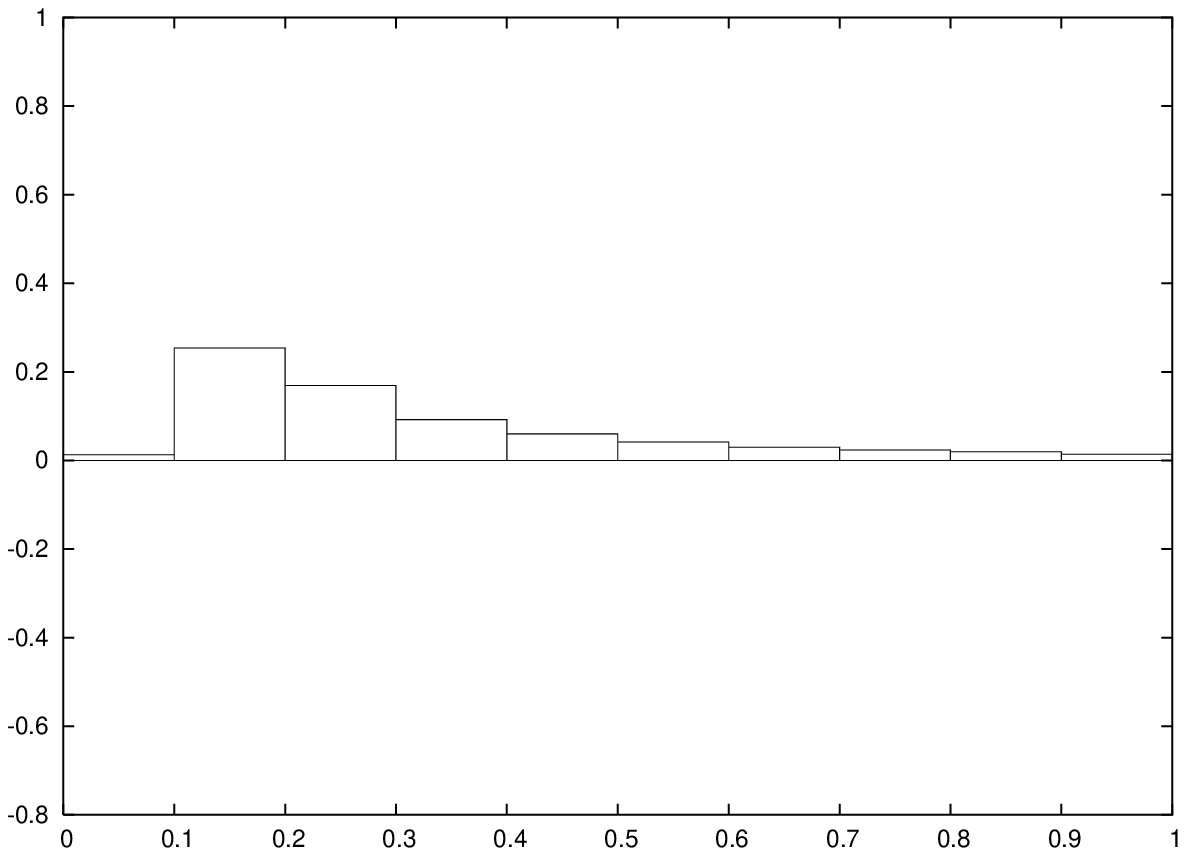}
\caption{The horizontal line is ${\alpha_s(t)/\pi}$ and the vertical
axis is the distribution from the each energy region 
in ${\alpha_s}$ to the total decay
 amplitude. 
The left figure is the real part and right one is
imaginary part
of the ${u}$ loop contribution.
These figures show that nonperturbative contributions
are important, so we cannot accurately compute the ${u}$ quark loop
contribution and have to take into account the
hadronic uncertainty. 
}
\label{u}
\end{figure}
\end{widetext}

\subsection{Annihilation contributions }
\subsubsection{Tree annihilation}
\begin{figure}[htbp]
\includegraphics[width=3.8cm]{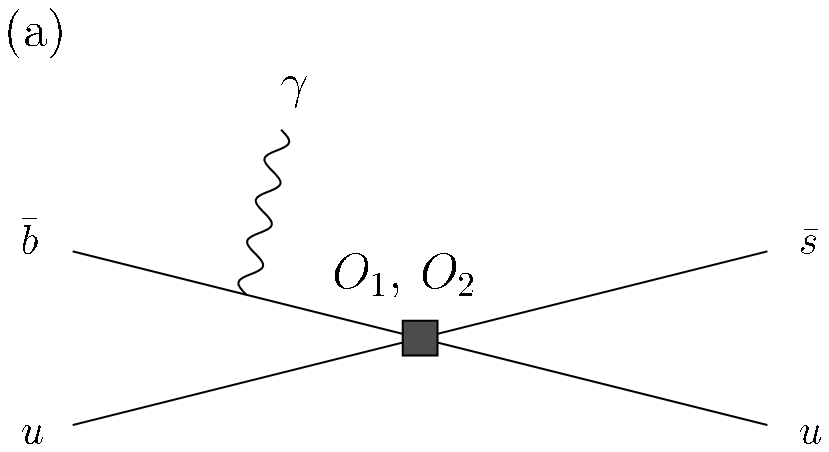}
\hspace{5mm}
\includegraphics[width=3.8cm]{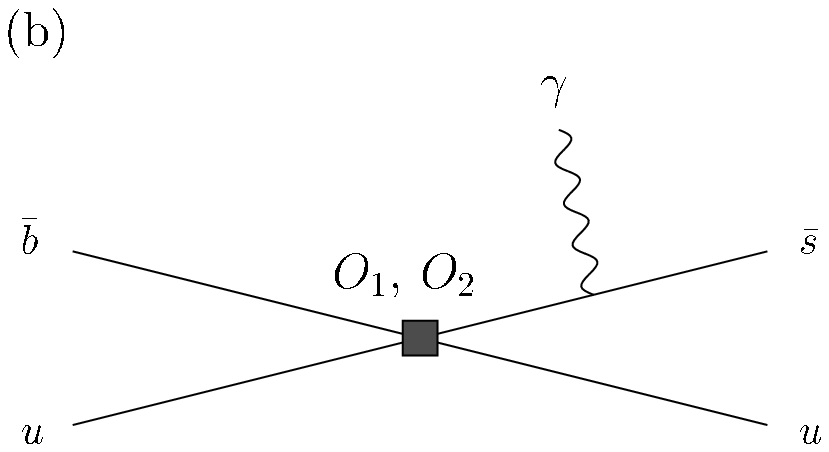}
\vspace{1cm}
\\
\includegraphics[width=3.8cm]{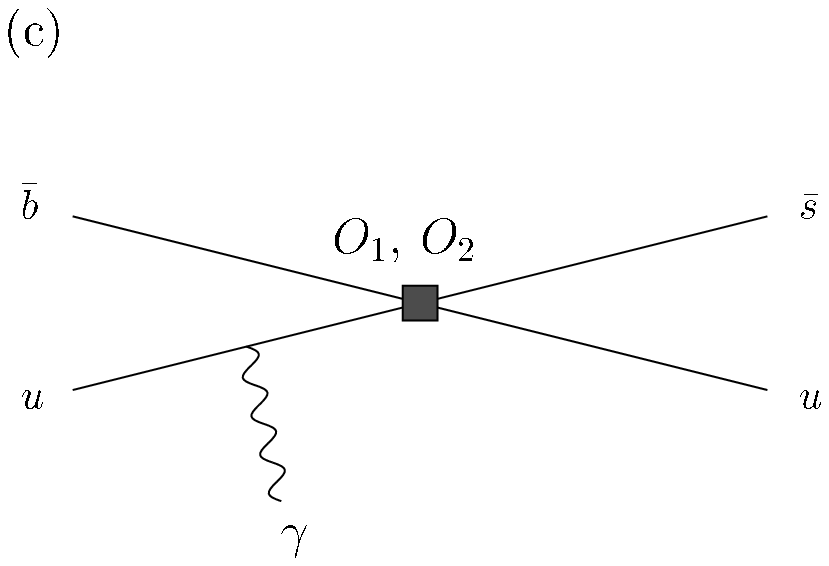}
\hspace{5mm}
\includegraphics[width=3.8cm]{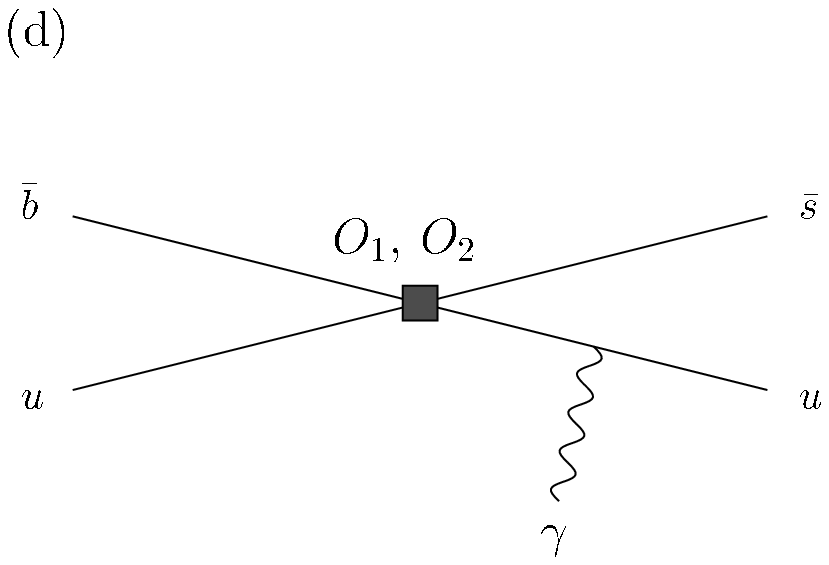}
\caption{Annihilation diagrams caused by  
 ${O_{1}}$, ${O_2}$ operators. In these cases,
no hard gluons are needed because they are
four Fermi interactions
and do not include spectator quarks which should be line up to
form hadrons. 
}
\label{annihilation}
\end{figure}

We now discuss  the annihilation contributions
caused by ${O_1}$  and ${O_2}$ operators.
The diagrams are shown in Fig.\ref{annihilation}.
The operators ${O_1}$, ${O_2}$ can be rewritten as
\small{
\begin{eqnarray}
O_1&=&(\bar{s}_i u_j)_{V-A}(\bar{u}_j b_i)_{V-A}=
(\bar{s}_i b_i)_{V-A}(\bar{u}_j u_j)_{V-A},\hspace{2mm}\\
O_2&=&(\bar{s}_i u_i)_{V-A}(\bar{u}_j b_j)_{V-A}=
(\bar{s}_i b_j)_{V-A}(\bar{u}_j u_i)_{V-A}.
\end{eqnarray}
}
These annihilation contributions are tree processes:
no hard gluons are needed because they are four Fermi interaction processes
and do not include spectator quarks which should be line up to
form hadrons. However, these contributions are small
because it has ${(V-A)\otimes(V-A)}$ vertex:
gets chiral suppression, and its' CKM factor is ${V_{ub}^*V_{us}}$,
${O(\lambda^2)}$ suppression compared to ${V_{tb}^*V_{ts}}$ and
${V_{cb}^*V_{cs}}$. Defining ${a_2(t)=C_2(t)+C_1(t)/3}$,
the each decay amplitudes are as follows:

\begin{eqnarray}
M_2^{S(a)}&=&M_2^{P(a)}\nonumber\\
&=&-F^{(0)}\xi_u\frac{3\sqrt{6}Q_bf_{K^*}\pi}{4M_B^2}r_{K^*}
\int^1_0 dx_1\int b_1 db_1 \nonumber\\
&&
a_2(t_a^a)S_t(x_1)
\hspace{1mm}
e^{\left[-S_B(t_a^a)\right]}\phi_B(x_1,b_1) K_0(b_1 A_a)
\nonumber\\
&&
\hspace{5mm}
\left(t_a^a=\mbox{max}(A_a,1/b_1)\right)
\end{eqnarray}
\begin{eqnarray}
M_2^{S(b)}
&=&-F^{(0)}\xi_u\frac{3\sqrt{6}Q_sf_B\pi}{4M_B^2}r_{K^*}
\int^1_0 dx_2 \int b_2 db_2\nonumber\\
&&a_2(t_a^b)S_t(x_2)\hspace{1mm}
e^{\left[-S_{K^*}(t_a^b)\right]}i\frac{\pi}{2}H_0^{(1)}(b_2B_a)\nonumber\\
&\times&\Big[(2-x_2)\phi_{K^*}^{v}(x_2)+x_2 \phi_{K^*}^a(x_2)\Big]
\end{eqnarray}
\begin{eqnarray}
M_2^{P(b)}
&=&F^{(0)}\xi_u\frac{3\sqrt{6}Q_sf_{B}\pi}{4M_B^2}r_{K^*}
\int^1_0 dx_2 \int b_2 db_2 \nonumber\\
&&a_2(t_a^b)S_t(x_2)\hspace{1mm}
e^{\left[-S_{K^*}(t_a^b)\right]}i\frac{\pi}{2}H_0^{(1)}(b_2B_a)\nonumber\\
&\times &\left[x_2\phi_{K^*}^{v}(x_2)+(2-x_2)\phi_{K^*}^a(x_2)\right]
\nonumber\\
&&
\hspace{5mm}\left(t_a^b=\mbox{max}(B_a,1/b_2)\right)
\end{eqnarray}
\begin{eqnarray}
M_2^{S(c)}&=&-M_2^{P(c)}\nonumber\\
&=&F^{(0)}\xi_u\frac{3\sqrt{6}Q_uf_{K^*}\pi}{4M_B^2}r_{K^*}
\int^1_0 dx_1\int b_1 db_1\nonumber\\
&&a_2(t_a^c)S_t(x_1)e^{\left[-S_B(t_a^c)\right]}\phi_B(x_1,b_1)
K_{0}(b_1C_a)
\nonumber\\
&&
\hspace{5mm}\left(t_a^c=\mbox{max}(C_a,1/b_1)\right)
\end{eqnarray}
\begin{eqnarray}
M_2^{S(d)}
&=&F^{(0)}\xi_u\frac{3\sqrt{6}Q_uf_B\pi}{4M_B^2}r_{K^*}
\int^1_0 dx_2 \int b_2 db_2\nonumber\\
&&a_2(t_a^d)S_t(x_2)\hspace{1mm}
e^{\left[-S_{K^*}(t_a^d)\right]}i\frac{\pi}{2}H_0^{(1)}(b_2D_a)
\nonumber\\
&\times &
\left[(1+x_2)\phi_{K^*}^{v}(x_2)-(1-x_2) \phi_{K^*}^a(x_2)\right]
\end{eqnarray}
\begin{eqnarray}
M_2^{P(d)}
&=&F^{(0)}\xi_u\frac{3\sqrt{6}Q_uf_{B}\pi}{4M_B^2}r_{K^*}
\int^1_0 dx_2 \int b_2 db_2\nonumber\\
&&a_2(t_a^d)S_t(x_2)\hspace{1mm}
e^{\left[-S_{K^*}(t_a^d)\right]}i\frac{\pi}{2}H_0^{(1)}(b_2D_a)\nonumber\\
&\times
 &\left[(1-x_2)\phi_{K^*}^{v}(x_2)-(1+x_2)\phi_{K^*}^a(x_2)\right]
\nonumber\\
&&
\hspace{5mm}\left(t_a^d=\mbox{max}(D_a,1/b_2)\right)
\end{eqnarray}
\begin{eqnarray}
&&A_a^2=(1+x_1)M_B^2,\hspace{1cm}B_a^2=(1-x_2)M_B^2,\nonumber\\
&&C_a^2=x_1M_B^2,
\hspace{1cm}D_a^2=x_2M_B^2
\end{eqnarray}

\subsubsection{QCD Penguin}
\begin{figure}[htbp]
\includegraphics[width=3.8cm]{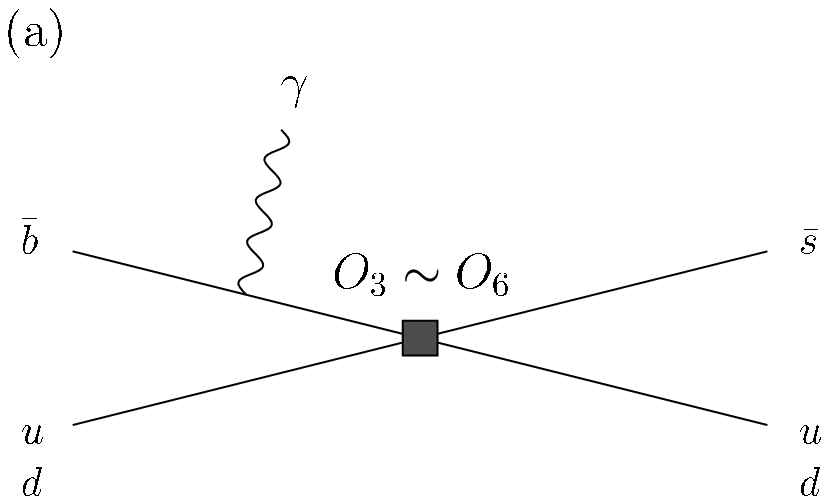}
\hspace{5mm}
\includegraphics[width=3.8cm]{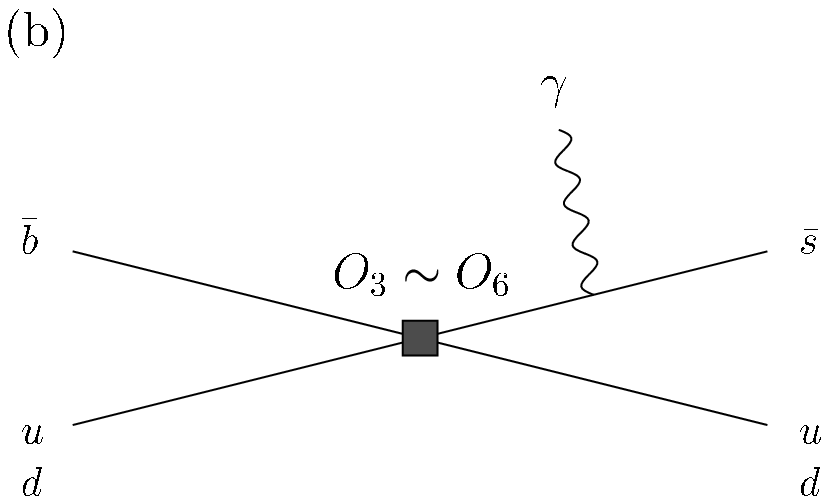}
\vspace{1cm}
\\
\includegraphics[width=3.8cm]{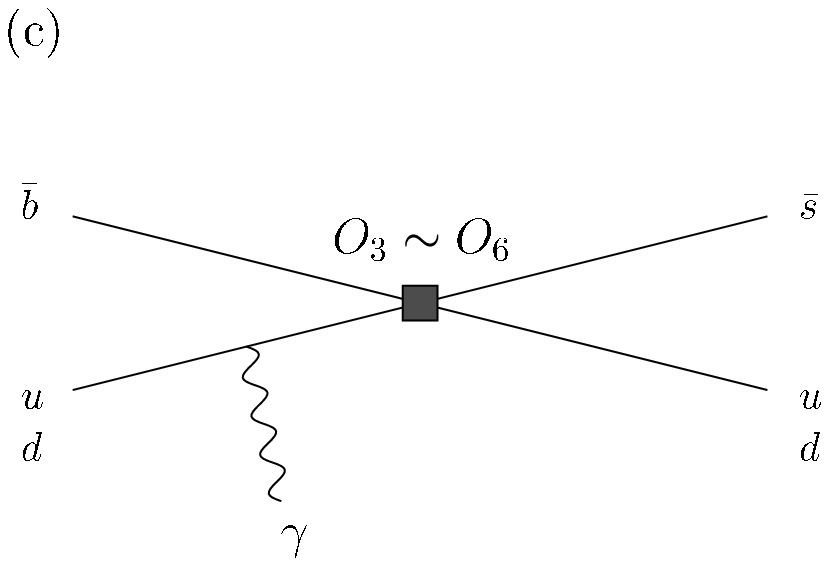}
\hspace{5mm}
\includegraphics[width=3.8cm]{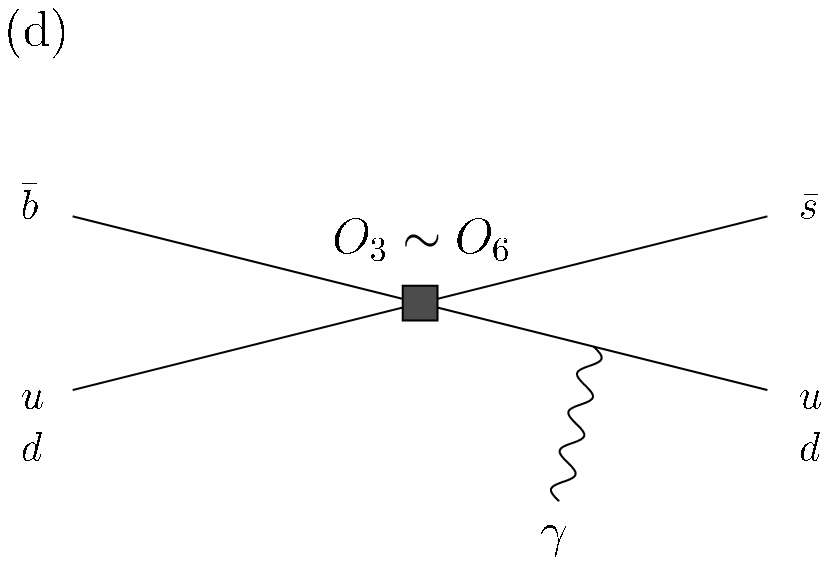}
\caption{Annihilation diagrams caused by  
 ${O_{3}\sim O_6}$ operators. 
QCD penguin annihilations include ${\alpha_s}$ in Wilson coefficient,
so they are the same order of all contributions
except for ${O_1}$, ${O_2}$ annihilations.
}
\label{penguin}
\end{figure}
Next we mention  the QCD penguin annihilation
caused by ${O_3\sim O_6}$ operators like in Fig.\ref{penguin}.
Here we define ${a_4(t)=C_4(t)+C_3(t)/3,
\hspace{2mm}a_6(t)=C_6(t)+C_5(t)/3}$.
${O_3}$, ${O_4}$ have the same expression
of ${O_1}$, ${O_2}$ annihilation contributions.
${O_5}$,  ${O_6}$ have a
${(V-A)\otimes(V+A)}$ vertex so they have chiral enhancement
compared to ${(V-A)\otimes(V-A)}$ vertex,
and its' CKM factor are ${V_{tb}^*V_{ts}}$,
then its' contributions are 
comparatively large
and get main origins for isospin breaking effects.
\begin{eqnarray}
M_4^{S(a)}&=&M_4^{P(a)}\nonumber\\
&=&F^{(0)}\xi_t\frac{3\sqrt{6}Q_bf_{K^*}\pi }{4M_B^2}r_{K^*}
\int^1_0 dx_1\int b_1 db_1 \nonumber\\
&&a_4(t_a^a)S_t(x_1)
\hspace{1mm}
e^{\left[-S_B(t_a^a)\right]}\phi_B(x_1,b_1) K_0(b_1 A_a)\nonumber\\
&&\hspace{5mm}
\left(t_a^a=\mbox{max}(A_a,1/b_1)\right)
\end{eqnarray}
\begin{eqnarray}
M_4^{S(b)}
&=&F^{(0)}\xi_t\frac{3\sqrt{6}Q_sf_B\pi}{4M_B^2}r_{K^*}
\int^1_0 dx_2 \int b_2 db_2\nonumber\\
&&a_4(t_a^b)S_t(x_2)\hspace{1mm}
e^{\left[-S_{K^*}(t_a^b)\right]}i\frac{\pi}{2}H_0^{(1)}(b_2B_a)\nonumber\\
&\times &
\Big[(2-x_2)\phi_{K^*}^{v}(x_2)+x_2 \phi_{K^*}^a(x_2)\Big]
\end{eqnarray}
\begin{eqnarray}
M_4^{P(b)}
&=&-F^{(0)}\xi_t\frac{3\sqrt{6}Q_sf_{B}\pi}{4M_B^2}r_{K^*}
\int^1_0 dx_2 \int b_2 db_2\nonumber\\
&&a_4(t_a^b)S_t(x_2)\hspace{1mm}
e^{\left[-S_{K^*}(t_a^b)\right]}i\frac{\pi}{2}H_0^{(1)}(b_2B_a)\nonumber\\
&\times &\left[x_2\phi_{K^*}^{v}(x_2)+(2-x_2)\phi_{K^*}^a(x_2)\right]
\nonumber\\
&&
\hspace{5mm}\left(t_a^b=\mbox{max}(B_a,1/b_2)\right)
\end{eqnarray}
\begin{eqnarray}
M_4^{S(c)}&=&-M_4^{P(c)}\nonumber\\
&=&-F^{(0)}\xi_t\frac{3\sqrt{6}Q_qf_{K^*}\pi}{4M_B^2}r_{K^*}
\int^1_0 dx_1\int b_1 db_1\nonumber\\
&&a_4(t_a^c)S_t(x_1)e^{\left[-S_B(t_a^c)\right]}\phi_B(x_1,b_1)
K_{0}(b_1C_a)\nonumber\\
&&
\hspace{5mm}\left(t_a^c=\mbox{max}(C_a,1/b_1)\right)
\end{eqnarray}
\begin{eqnarray}
M_4^{S(d)}
&=&-F^{(0)}\xi_t\frac{3\sqrt{6}Q_qf_B\pi}{4M_B^2}r_{K^*}
\int^1_0 dx_2 \int b_2 db_2\nonumber\\
&&a_4(t_a^d)S_t(x_2)\hspace{1mm}
e^{\left[-S_{K^*}(t_a^d)\right]}i\frac{\pi}{2}H_0^{(1)}(b_2D_a)\nonumber\\
&\times &\left[(1+x_2)\phi_{K^*}^{v}(x_2)-(1-x_2)
			\phi_{K^*}^a(x_2)\right]
\end{eqnarray}
\begin{eqnarray}
M_4^{P(d)}
&=&-F^{(0)}\xi_t\frac{3\sqrt{6}Q_qf_{B}\pi}{4M_B^2}r_{K^*}
\int^1_0 dx_2 \int b_2 db_2\nonumber\\
&&a_4(t_a^d)S_t(x_2)\hspace{1mm}
e^{\left[-S_{K^*}(t_a^d)\right]}i\frac{\pi}{2}H_0^{(1)}(b_2D_a)\nonumber\\
&\times
 &\left[(1-x_2)\phi_{K^*}^{v}(x_2)-(1+x_2)\phi_{K^*}^a(x_2)\right]
\nonumber\\
&&
\hspace{5mm}\left(t_a^d=\mbox{max}(D_a,1/b_2)\right)
\end{eqnarray}
\begin{eqnarray}
M_6^{S(b)}&=&-M_6^{P(b)}\nonumber\\
&=&-F^{(0)}\xi_t\frac{3\sqrt{6}Q_s f_B\pi}{2M_B^2}
\int^1_0 dx_2 \int b_2 db_2 ~a_6(t_a^b)\nonumber\\
&\times &S_t(x_2)e^{-\left[S_{K^*}(t_a^b)\right]}
\phi_{K^*}^T(x_2) i\frac{\pi}{2}H_0^{(1)}(b_2 B_a)\nonumber\\
&&\hspace{5mm}\left(t_a^b=\mbox{max}(B_a,1/b_2)\right)
\end{eqnarray}
\begin{eqnarray}
M_6^{S(d)}&=&-M_6^{P(d)}\nonumber\\
&=&-F^{(0)}\xi_t\frac{3\sqrt{6}Q_q f_B\pi}{2M_B^2}
\int^1_0 dx_2 \int b_2 db_2 ~ a_6(t_a^d)\nonumber\\
&\times &S_t(x_2)e^{-\left[S_{K^*}(t_a^d)\right]}
\phi_{K^*}^T(x_2) i\frac{\pi}{2}H_0^{(1)}(b_2 D_a)\nonumber\\
&&
\hspace{5mm}\left(t_a^d=\mbox{max}(D_a,1/b_2)\right)
\end{eqnarray}
\begin{eqnarray}
&&A_a^2=(1+x_1)M_B^2,\hspace{5mm}B_a^2=(1-x_2)M_B^2,\nonumber\\
&&C_a^2=x_1M_B^2,
\hspace{1cm}D_a^2=x_2M_B^2
\end{eqnarray}

\section{Long distance contributions to the photon quark coupling}
\label{Long distance}
\begin{figure}[htbp]
\includegraphics[width=5cm]{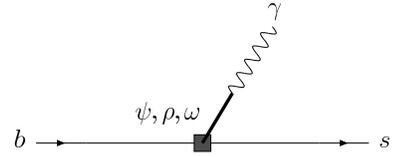}
\caption{Vector-Meson-Dominance contributions
mediated by ${\psi,\rho,\omega}$.}
\label{Vector}
\end{figure}

Here we want to discuss the long distance contributions.
In order to examine the standard model or search for new physics
indirectly by comparing the experimental data with the
values predicted within the standard model, 
we have to take into account these long distance effects:
 ${B\to K^* (J/\psi,\rho,\omega) \rightarrow K^* \gamma}$
\cite{Golowich:1994zr,Deshpande:1994cn} (Fig.\ref{Vector}).
It should be noted that ${B\to D\bar{D}K^*\to K^*\gamma}$
is small compared to the ${J/\psi}$ intermediate state contribution.

\begin{figure}
\includegraphics[width=3.8cm]{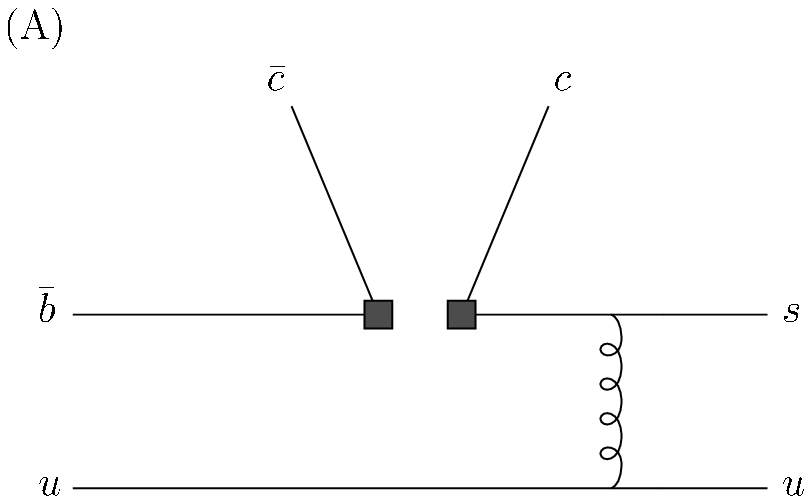}
\hspace{5mm}
\includegraphics[width=3.8cm]{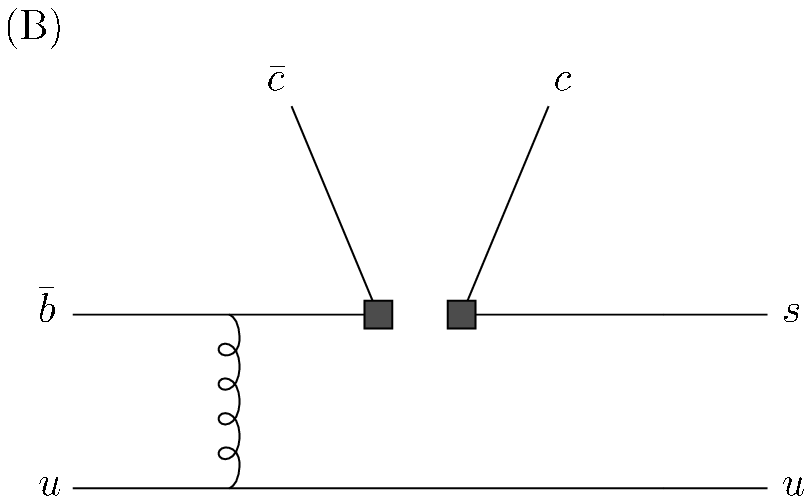}
\vspace{3mm}
\\
\includegraphics[width=3.8cm]{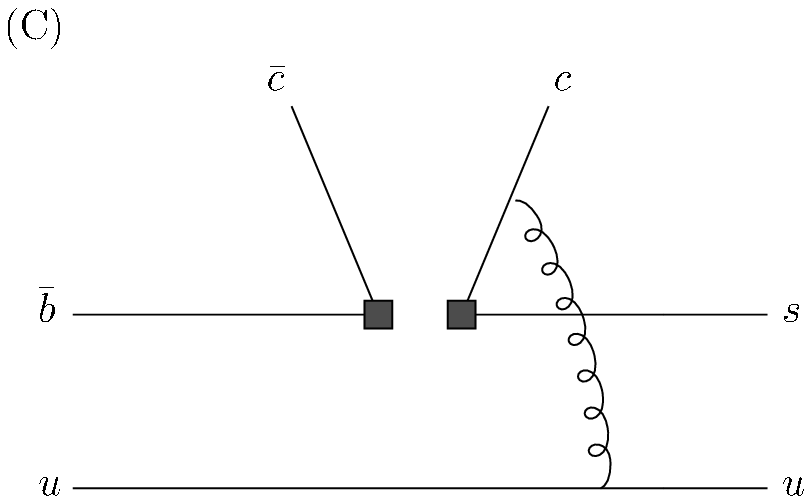}
\hspace{5mm}
\includegraphics[width=3.8cm]{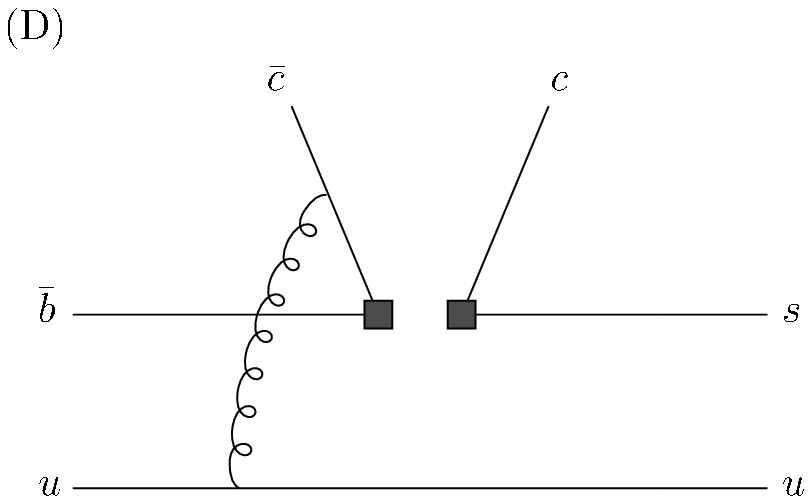}
\caption{(A), (B) are factorizable and (C), (D) are nonfactorizable
 contributions to the hadronic matrix element for ${<K^*\psi |H_{eff}|B>}$. }
\label{non-fact}
\end{figure}

These contributions are caused by ${O_1}$, ${O_2}$ operators,
and the effective Hamiltonian describing these processes is
\begin{eqnarray}
H_{eff}=\frac{G_F}{\sqrt 2}\sum_{q=u,c} V_{qb}V_{qs}^*(C_1(t)O_1^{(q)}(t)
+ C_2(t)O_2^{(q)}(t))+\mbox{h.c.}\nonumber\\
\end{eqnarray}
If we use the vector-meson-dominance,
the ${B\to K^*\gamma}$ decay amplitude can be expressed
as inserting the complete set of possible intermediate
vector meson states like
\begin{eqnarray}
 \langle K^*\gamma | H_{eff}|B\rangle
=\sum_{V}\langle\gamma | A_{\nu}J_{em}^{\nu}|V\rangle
\frac{-i}{q_V^2-m_V^2}
\langle VK^*  | H_{eff}|B\rangle,
\nonumber\\
\end{eqnarray}
where ${V=\psi,\rho,\omega}$.
Now we concretely consider the ${B\rightarrow K^* \psi \to K^*\gamma}$.
Four diagrams contribute to the
hadronic matrix element of
${\langle K^* \psi  | H_{eff}|B\rangle}$ (see Fig.\ref{non-fact}),
and first of all, 
we consider the leading contributions:
the factorizable ones, Figs. \ref{non-fact}(A) and \ref{non-fact}(B).
\subsection{Factorizable contribution}
In this case,
the ${B\to \psi K^*}$ decay amplitude can be factorized 
as
\begin{eqnarray}
\langle \psi K^* \mid H_{eff}\mid B\rangle
&=&\frac{G_F}{\sqrt{2}}V_{cb}V_{cs}^* a_1(t)
\langle \psi \mid \bar{c}\gamma_{\mu}(1-\gamma^5)c\mid
0\rangle\nonumber\\
&&\times 
\langle K^*\mid \bar{s}\gamma^{\mu}(1-\gamma^5)b\mid B\rangle,
\label{62}
\end{eqnarray}
and the definition of the decay constant is
\begin{eqnarray}
\langle \psi (q)\mid \bar c \gamma_{\mu}c \mid 0 \rangle
\equiv  i m_{\psi}g_{\psi}(q^2)\epsilon_{\psi \mu}^{*}(q),
\end{eqnarray}
then the decay amplitude can be written as
\begin{eqnarray}
\label{amplitude-1}
M(B\rightarrow K^* \psi (q))&=&\frac{G_F}{\sqrt 2}V_{cb}V_{cs}^*
 a_1(t)im_{\psi}g_{\psi}(q^2)\epsilon^{*}_{\psi \mu}(q)\nonumber\\
&&\times 
 \langle K^*\mid 
\bar s \gamma^{\mu}(1-\gamma^5)b \mid B\rangle ,
\end{eqnarray}
where ${a_1(t)=C_1(t)+C_2(t)/3}$. 
The conversion part of the ${\psi}$ meson into photon can
be expressed as
\begin{eqnarray}
\langle \gamma \mid A_{\nu}J_{em}^{\nu}\mid\psi\rangle
=-\frac{2}{3}em_{\psi}g_{\psi}(q^2),
\end{eqnarray}
then the total amplitude of ${B\to K^*\gamma}$ mediated
by ${\psi}$ meson can be expressed as follows,
\begin{eqnarray}
M(B\rightarrow K^* \psi \to K^*\gamma)&=&\frac{G_F}{\sqrt{2}}
V_{cb}V_{cs}^* a_1(t)\left(\frac{2eg_{\psi}(0)^2}{3}\right)
\epsilon_{\psi\mu}^*\nonumber\\
&&\times 
\langle K^*\mid \bar{s}\gamma^{\mu}(1-\gamma^5)b\mid B\rangle
\end{eqnarray}
where the real photon momentum is
${q^2=0}$.
In principle, we need to include the width of the vector meson 
in the propagator and write
\begin{eqnarray}
\frac{-i}{q^2 -m^2 +im\Gamma}~,
\end{eqnarray}
but
we have 
${(\Gamma_{\psi}/m_{\psi})\sim O(10^{-5})}$
and the effects of the width can be safely neglected.

The amplitudes for
${B\rightarrow K^* \omega \rightarrow K^* \gamma}$ can be 
computed in a similar manner.
In this case, ${(\Gamma_{\omega}/m_{\omega})=1.0\times 10^{-2}}$
and we can also neglect the width effect in the 
meson propagator.
Differences with ${B\rightarrow K^* \psi \rightarrow K^* \gamma}$ 
are the value of decay constant ${g_{\omega}(0)}$
and the factor for the electromagnetic interaction.
\begin{eqnarray}
M(B\rightarrow K^* \omega\rightarrow K^* \gamma )
&=&\frac{G_F}{\sqrt 2 }V_{ub}V_{us}^*a_1(t)
\left(\frac{eg_{\omega}(0)^2}{6}\right)\epsilon_{\omega\mu}^*\nonumber\\
&&\times 
\langle K^* \mid\bar s \gamma^{\mu}(1-\gamma^5)b\mid B\rangle
\end{eqnarray} 
However in the ${B\rightarrow K^* \rho\rightarrow K^* \gamma }$ case,
the ${\rho}$ resonance peak is not so sharp,
so the propagation of ${\rho}$ meson generates
the strong phase:
${(\Gamma_{\rho}/m_{\rho})\simeq 0.19}$ and it introduces
${\simeq 11 ^{\circ}}$ strong phase.
\begin{eqnarray}
M(B\rightarrow K^* \rho\rightarrow K^* \gamma )
&=&\frac{G_F}{\sqrt 2 }V_{ub}V_{us}^*a_1(t)
\left(\frac{eg_{\rho}(0)^2}{2
 (1-i\Gamma_{\rho}/m_{\rho})}
\right)\epsilon_{\rho\mu}^*\nonumber\\
&&\times 
\langle K^*\mid \bar s \gamma^{\mu}(1-\gamma^5)b\mid B\rangle
\end{eqnarray} 
In order to estimate these long distance contributions,
we have to know the decay constant ${g_V}$.
The decay constants are experimentally determined
by the ${V\rightarrow e^+ e^-}$ data \cite{Eidelman:2004wy}.
The amplitude for ${V\to e^+e^-}$ can be expressed as
\begin{eqnarray}
M(V\to e^+e^-)=Qe^2m_{V}g_V(q^2)
\end{eqnarray}
where ${Q}$ expresses the electric charge like that
${Q=Q_c}$ when ${V=\psi}$, ${Q=(Q_u-Q_d)/\sqrt{2}}$
in ${V=\rho}$ case, and in ${V={\omega}}$ case,
${Q=(Q_u+Q_d)/\sqrt{2}}$. 
Then the decay width for ${V\rightarrow e^+ e^-}$ decay
can be written like
\begin{eqnarray}
\Gamma(V\rightarrow e^+ e^-)=\frac{4\pi Q^2\alpha_{em}^2g_V^2(q^2)}{3m_V}~,
\end{eqnarray}
and the values of ${g_V}$ are in Table \ref{constant}.
\begin{center}
\begin{table}[htbp]
\begin{tabular}{c c c c}
\hline
\hline
${V}$ & ${\Gamma(V\rightarrow e^+ e^-)(GeV)}$&
 ${m_V(GeV)}$&${g_V^2(GeV^{2})}$\\
\hline
${J/\psi(1S)}$&${5.26\times 10^{-6}}$& 3.097&0.1642\\
\hline
${\psi(2S)}$&${2.19\times 10^{-6}}$& 3.686&0.0814\\
\hline
${\psi(3770)}$&${0.26\times 10^{-6}}$&3.770 &0.0099\\
\hline
${\psi(4040)}$&${0.75\times 10^{-6}}$&4.040 &0.0306\\
\hline
${\psi(4160)}$&${0.77\times 10^{-6}}$& 4.160&0.0323\\
\hline
${\psi(4415)}$&${0.47\times 10^{-6}}$&4.415 &0.0209\\
\hline
${\rho}$&${7.02\times 10^{-6}}$&0.771&0.0485\\
\hline
${\omega}$&${0.60\times 10^{-6}}$&0.783&0.0379\\
\hline
\end{tabular}
\caption{The coefficients ${g_V}$.}
\label{constant}
\end{table}
\end{center}

\begin{widetext}
Furthermore, these decay constants are defined at the 
${q^2=m_V^2}$ energy scale.
We need ones at ${q^2=0}$,
so we have to extrapolate
these decay constants from ${q^2=m_V^2}$
to ${q^2=0}$.
We express ${g_V(0)}$ 
as ${g_V(0)=\kappa g_V(q^2)}$ by using suppression 
factor ${\kappa}$.
In the ${\psi}$ cases, we take ${\kappa\simeq 0.4}$ 
\cite{Golowich:1994zr,Deshpande:1994cn},
and in the 
${\rho,\omega}$ cases, we take ${\kappa \simeq 1.0}$
\cite{Anderson:1976hi,Paul:1981dw}.
Then the long distance contributions mediated by ${\psi,\rho,\omega}$
are 
\begin{eqnarray}
M(B\rightarrow K^* \gamma )
&=&\frac{G_F}{\sqrt 2 }a_1(t)e\Big(V_{cb}V_{cs}^*
\frac{2\kappa g_{\psi}(m_\psi^2)^2}{3}
+V_{ub}V_{us}^*\left[
\frac{g_{\omega}(m_{\omega}^2)^2}{6}+
\frac{g_{\rho}(m_{\rho}^2)^2}{2
 (1-i\Gamma/m_{\rho})}
\right]\Big)
\epsilon_{\gamma\mu}
\langle K^*\mid \bar s \gamma^{\mu}(1-\gamma^5)b\mid B\rangle,
\end{eqnarray}
and if we calculate the form factor of ${\langle K^*\mid \bar s
\gamma^{\mu}(1-\gamma^5)b\mid B\rangle}$,
the long distance contributions become as follows:

\begin{eqnarray}
M^{S(A)}&=&-M^{P(A)}\nonumber\\
&=&\frac{8\pi^2}{M_B^2}F^{(0)}\int^1_0 dx_1 dx_2
\int b_1db_1 b_2db_2\phi_B(x_1,b_1)S_t(x_2)\alpha_s(t_7^a)
e^{[-S_B(t_7^a)-S_{K^*}(t_7^a)]}
a_1(t_7^a)
H_{7}^{(a)}(A_7b_2,B_7b_1, B_7b_2)\nonumber\\
&&\times r_{K^*}\left[\phi_{K^*}^v(x_2)+\phi_{K^*}^a(x_2)\right]
\left(
\xi_{c}\frac{2\kappa g_{\psi}(m_\psi^2)^2}{3}
+
\xi_{u}\left[
\frac{g_{\omega}(m_{\omega}^2)^2}{6}+
\frac{g_{\rho}(m_{\rho}^2)^2}{2
 (1-i\Gamma/m_{\rho})}
\right]\right)
\nonumber\\
&&
\hspace{3cm}
\left(t_7^a=max(A_7,B_7,1/b_1,1/b_2)\right)
\end{eqnarray}
\begin{eqnarray}
M^{S(B)}&=&\frac{8\pi^2}{M_B^2}F^{(0)}\int^1_0 dx_1 dx_2
\int b_1db_1 b_2db_2 )\phi_B(x_1,b_1)
S_t(x_2)
\alpha_s(t_7^b)
e^{[-S_B(t_7^b)-S_{K^*}(t_7^b)]}
a_1(t_7^b)
H_{7}^{(b)}(A_7b_1,C_7b_1, C_7b_2)\nonumber\\
&&\times \Big[(x_2+2)r_{K^*}\phi_{K^*}^v(x_2)
+\phi_{K^*}^T(x_2)
-x_2r_{K^*}\phi_{K^*}^a(x_2)\Big]
\left(
\xi_{c}\frac{2\kappa g_{\psi}(m_\psi^2)^2}{3}+
\xi_{u}\left[
\frac{g_{\omega}(m_{\omega}^2)^2}{6}+
\frac{g_{\rho}(m_{\rho}^2)^2}{2
 (1-i\Gamma/m_{\rho})}
\right]\right)\nonumber\\
\end{eqnarray}
\begin{eqnarray}
M^{P(B)}&=&-\frac{8\pi^2}{M_B^2}F^{(0)}\int^1_0 dx_1 dx_2
\int b_1db_1 b_2db_2 \phi_B(x_1,b_1)
S_t(x_2)\alpha_s(t_7^b)
e^{[-S_B(t_7^b)-S_{K^*}(t_7^b)]}
a_1(t_7^b)H_{7}^{(b)}(A_7b_1,C_7b_1, C_7b_2)\nonumber\\
&&\times \Big[-x_2r_{K^*}\phi_{K^*}^v(x_2)
+\phi_{K^*}^T(x_2)+(x_2+2)r_{K^*}\phi_{K^*}^a(x_2)\Big]
\left(
\xi_c\frac{2\kappa g_{\psi}(m_\psi^2)^2}{3}+
\xi_{u}\left[
\frac{g_{\omega}(m_{\omega}^2)^2}{6}+
\frac{g_{\rho}(m_{\rho}^2)^2}{2
 (1-i\Gamma/m_{\rho})}
\right]\right)\nonumber\\
&&\hspace{3cm}\left(t_7^b=max(A_7,C_7,1/b_1,1/b_2)\right)
\end{eqnarray}
\end{widetext}

\subsection{Nonfactorizable contribution}
Next we estimate the the effect of
nonfactorizable contributions  to 
the physical quantity like branching ratio,
CP asymmetry, and isospin breaking effects.
In order to do so in the case of ${B\to K^* \psi\to K^*\gamma}$ 
at first, we use the experimental data on the
branching ratio and different helicity amplitudes
for ${B\to J/\psi~K^* }$ decay mode.
The branching ratio is 
${Br(B^0\to J/\psi ~K^{*0})=(1.31\pm 0.07)\times
10^{-3}}$\cite{Eidelman:2004wy},
and 
the fraction of the
transversely polarized decay width
to the total decay width is about
${\Gamma_T/\Gamma=\simeq 0.4}$
\cite{Abe:2002ha,Aubert:2001pe,Jpsi},
then the corresponding 
transversally polarized
branching ratio
amounts to 
\begin{eqnarray}
Br(B\to J/\psi~ K^*)_T\simeq 5.0 \times 10^{-4}.
\label{77}
\end{eqnarray}
On the other hand, if we compute the branching ratio
by using eq.(\ref{62}),
we have
\begin{eqnarray}
Br(B\to J/\psi ~K^*)_T\simeq
2.3\times 10^{-4}.
\label{78}
\end{eqnarray}
If we assume that the difference between the
experimental value eq.(\ref{77})
and our prediction eq.(\ref{78})
is due to the nonfactorizable amplitude,
then
\begin{eqnarray}
\frac{\mbox{nonfactorizable}~A(B\to J/\psi~K^*)_T}
{\mbox{factorizable}~A(B\to J/\psi~K^*)_T}\simeq 0.4~.
\end{eqnarray}
Note however, that ${B\to K^*\gamma}$
is dominated by the short distance amplitudes.
The long distance correction from
the factorizable diagram is about 4\%
of the total decay amplitude.
So when we add the nonfactorizable amplitude,
the long distance correction increases to 6\%
in the total amplitude, and 12\% in the branching ratio.
We have included these corrections
in our numerical estimates given below.

Furthermore, we estimate the effect of nonfactorizable
contribution to the direct CP asymmetry.
In general, a nonfactorized amplitudes
has a relative strong phase compared to
the factorized amplitude.
We already know that the 
nonfactorizable diagram amounts to about 2\%
to the short distance amplitude, then
we can numerically estimate the CP asymmetry uncertainty 
from the nonfactorizable diagram 
by introducing the strong phase as a free parameter.
We conclude that only less than 10\%
uncertainty is generated by the long distance
nonfactorizable amplitude, and as we will see later,
this error is small compared to the total uncertainty
in CP asymmetry
from other origins.
Finally, we mention that these long distance contributions
do not generate the isospin breaking effect,
the nonfactorizable contribution can be neglected
in computing the isospin breaking effects.

In the case of ${B\to  K^*(\rho,\omega)\to K^*\gamma}$,
we can expect that
the factorized amplitudes are dominant to the
total decay amplitude in ${B\to K^*(\rho,\omega)}$
by the analogy of ${B\to \rho\rho}$ decay \cite{Kagan:2004uw},
then we can neglect the nonfactorized contribution
to the above physical quantities.

\subsection{Another diagrams for long distance contributions
to the photon quark coupling}
\begin{figure}
\includegraphics[width=3.8cm]{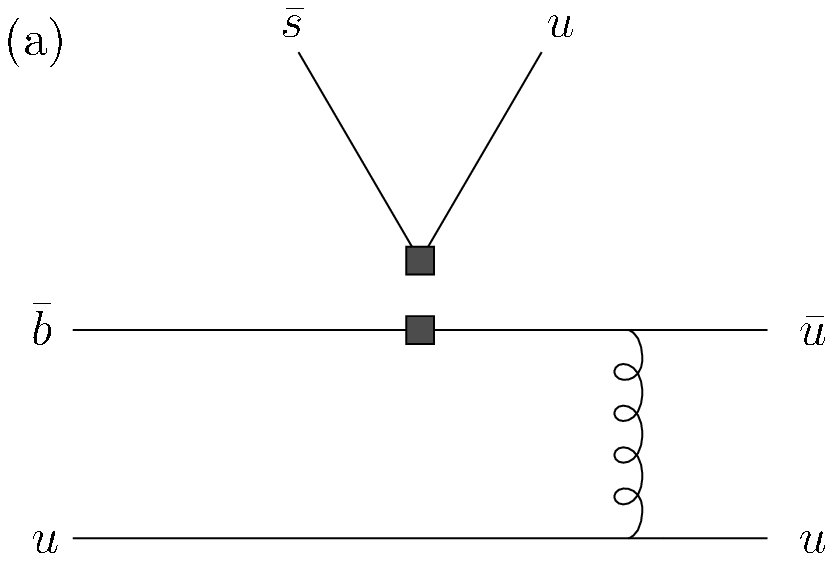}
\hspace{8mm}
\includegraphics[width=3.8cm]{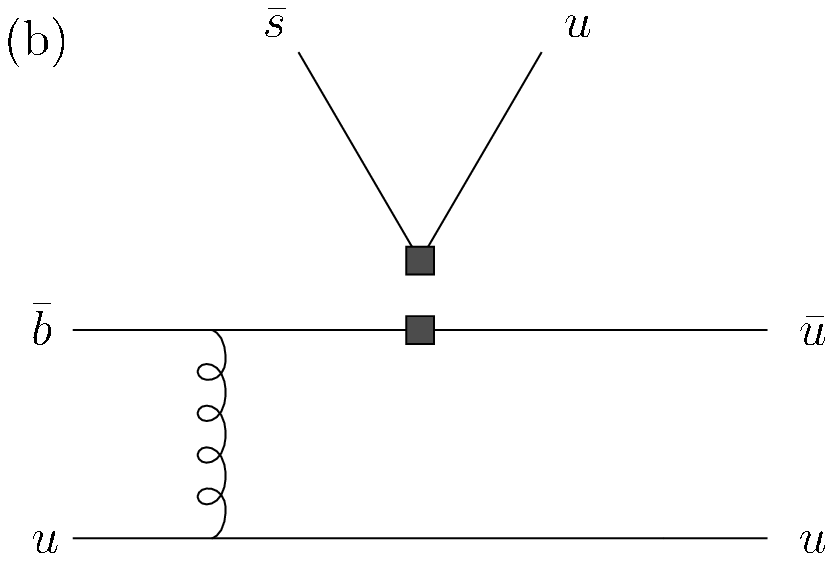}
\caption{Long distance effects mediated by ${\rho,\omega}$
which contribute only to the charged mode.}
\label{Long}
\end{figure}

Next we want to consider another contribution with different 
topology
which exist only
in the charged decay mode
like ${B^{\pm}\to K^{*\pm}\gamma}$ (Fig.\ref{Long}).
If we neglect the nonfactorizable contributions and annihilation
contributions,
there are two diagrams that contribute to
the hadronic matrix elements 
${\langle K^{*\pm }\rho(\omega) \mid H_{eff}\mid B^{\pm}\rangle}$.
We define the ${\rho}$ or ${\omega}$ meson momentum 
${P_3=M_B/\sqrt{2}\hspace{1mm}(1,0,\vec{0}_T)}$
and the spectator quark momentum 
fraction as ${x_3}$. 

\begin{widetext}
\begin{eqnarray}
M^{S(a)}&=&-M^{P(a)}\nonumber\\
&=&
\frac{4\pi^2f_{K^*}}{M_B^2}F^{(0)}\xi_{u}
\int^1_0 dx_1 dx_3 \int b_1 db_1 b_3 db_3~\phi_B(x_1,b_1)
\Big[\frac{g_{\rho}(m_{\rho}^2)}{1-i\Gamma/m_{\rho}}+\frac{g_{\omega}(m_{\omega}^2)}{3}\Big]S_t(x_1)\alpha_s(t^a)
e^{\left[-S_B(t^a)-S_{\rho}(t^a)\right]}\nonumber\\
&\times &a_2(t^a)r_{\rho}\left[
\phi_{\rho}^v(x_3)+\phi_{\rho}^a(x_3)
\right]H_7^{(a)}(A_7b_3,B_7b_1,B_7b_3)
\hspace{5mm}\left(t^a=\mbox{max}(A_7,B_7,1/b_1,1/b_3)\right)
\end{eqnarray}
\begin{eqnarray}
M^{S(b)}&=&
\frac{4\pi^2f_{K^*}}{M_B^2}F^{(0)}\xi_{u}
\int^1_0 dx_1 dx_3 \int b_1 db_1 b_3 db_3~\phi_B(x_1,b_1)
\Big[\frac{g_{\rho}(m_{\rho}^2)}{1-i\Gamma/m_{\rho}}+\frac{g_{\omega}(m_{\omega}^2)}{3}\Big]S_t(x_2)\alpha_s(t^b)
e^{\left[-S_B(t^b)-S_{\rho}(t^b)\right]}\nonumber\\
&\times &a_2(t^b)\left[
(x_3+2)r_{\rho}\phi_{\rho}^v(x_3)+\phi_{\rho}^T(x_3)-x_3r_{\rho}\phi_{\rho}^a(x_3)
\right]
H_7^{(b)}(A_7b_1,C_7b_1,C_7b_3)
\end{eqnarray}
\begin{eqnarray}
M^{P(b)}&=&-
\frac{4\pi^2f_{K^*}}{M_B^2}F^{(0)}\xi_{u}
\int^1_0 dx_1 dx_3 \int b_1 db_1 b_3 db_3~\phi_B(x_1,b_1)
\Big[\frac{g_{\rho}(m_{\rho}^2)}{1-i\Gamma/m_{\rho}}+\frac{g_{\omega}(m_{\omega}^2)}{3}\Big]S_t(x_2)\alpha_s(t^b)
e^{\left[-S_B(t^b)-S_{\rho}(t^b)\right]} \nonumber\\
&\times &a_2(t^b)\left[
-x_3r_{\rho}\phi_{\rho}^v(x_3)+\phi_{\rho}^T(x_3)
+(x_3+2)r_{\rho}\phi_{\rho}^a(x_3)
\right]
H_7^{(b)}(A_7b_1,C_7b_1,C_7b_3)\nonumber\\
&&
\hspace{3cm}\left(t^b= \mbox{max}(A_7,C_7,1/b_1,1/b_3)\right)
\end{eqnarray}
\begin{eqnarray}
A_7^2=x_1x_3M_B^2,\hspace{6mm}B_7^2=x_1M_B^2,
\hspace{6mm}C_7^2=x_3M_B^2
\end{eqnarray}

In the computation of the above formulas,
we use the ${\rho}$ and ${\omega}$ meson wave function
extracted from 
light-cone QCD sum rule \cite{Ball:1998ff},
and the detailed expression is in Appendix \ref{Appendix. B}.

\end{widetext}

\section{Numerical results}
\label{Numerical}
We want to show the numerical analysis in this section.
In the evaluation of the various form factors and amplitudes,
we adopt
${G_F=1.16639\times 10^{-5} \mbox{GeV}^{-2}}$,
leading order strong coupling ${\alpha_s}$ defined at 
the flavor number ${n_f=4}$, the decay constants 
${f_B=190 \mbox{MeV}}$, ${f_{K^*}=226 \mbox{MeV}}$,
and ${f_{K^*}^T=185 \mbox{MeV}}$,
the masses 
${M_B=5.28 \mbox{GeV}}$, ${M_{K^*}=0.892 \mbox{GeV}}$ and 
${m_c=1.2 \mbox{GeV}}$,
the meson lifetime ${\tau_{B^0}=1.542 \hspace{1mm}\mbox{ps}}$
and ${\tau_{B^+}=1.674\hspace{1mm}\mbox{ps}}$.
Furthermore we used the leading order Wilson coefficients 
\cite{Buchalla:1995vs} and we take the 
${K^*}$, ${\rho}$, and ${\omega}$ meson wave functions up to twist-3.
In order to make clear the theoretical error of the predicted physical
quantities, we want to show how to estimate these errors.
\subsection{Error Estimation}
\label{error}
When we estimate the physical quantities like
branching ratio, CP asymmetry, and isospin breaking effect,
there are four major classes of error in pQCD computations:
(1) the input parameter uncertainties;
(2) higher order effects in perturbation expansion;
(3) the CKM parameter uncertainties;
and (4) the hadronic uncertainties
from the ${u}$ quark loop.

\begin{widetext}
\begin{enumerate}
\item
First we want to estimate the class(1) error for
various physical quantities.
For class(1),
we change the decay constants,
the B meson wave function parameter
${\omega_B}$, and ${c}$ parameter of the threshold function.
We estimate the uncertainties from
the decay constants to be 15\% in the amplitude.
If we change the ${\omega_B}$ 
in the range ${\omega_B=(0.40\pm 0.04)}$GeV,
and ${c}$ in the range
${c=0.4\pm 0.1}$, 
these uncertainties
change the ${B\to K^*}$ form factor
by about 15\% at the amplitude level.
Thus we regard the total uncertainty for class(1)
to be 20\%.
Here we discuss how this error affects 
the experimental observables such as
the branching ratio, direct CP asymmetry,
and isospin breaking.

\begin{itemize}
\item \underline{Branching Ratio}

In order to see how much error is generated
when we change some parameters in class(1),
we introduce real parameter 
${\delta_i^j}$'s
as the fractional differences of the amplitudes from ones with
a fixed hadronic parameter, 
where ${i}$ and ${j}$ express
the flavor and electric charge.
Note that the uncertainty in decay constants 
leads to an uncertainty in overall factor
of the amplitude, i.e.
they don't lead to an uncertainty in the phase
of the amplitude.
In the change wave function parameters on the other hand,
the phase changes a little, but it's effect is very small
and we can introduce ${\delta_i^j}$'s as real parameters.

The decay widths of the ${B}$ and ${\bar{B}}$ meson
decays can be expressed as
\begin{eqnarray}
\Gamma({B}^j)&=&|V_{tb}^*V_{ts}A_t^j(1+\delta_t^j)
+V_{cb}^*V_{cs}A_c^j(1+\delta_c^j)
+V_{ub}^*V_{us}A_u^j(1+\delta_u^j)|^2
\label{81}\\
\Gamma(\bar{B}^j)&=&|V_{tb}V_{ts}^*A_t^j(1+\delta_t^j)
+V_{cb}V_{cs}^*A_c^j(1+\delta_c^j)
+V_{ub}V_{us}^*A_u^j(1+\delta_u^j)|^2,
\label{82}
\end{eqnarray}
and we can see that the uncertainty to the
branching ratio from input parameters
comes from the error of the ${O_{7\gamma}}$ amplitude,
and it amounts to
about ${2\delta_t^j\simeq 40\%}$.
\item \underline{Direct CP Asymmetry}

From eq.(\ref{81}) and (\ref{82}),
the direct CP asymmetry can be expressed as follows,
\begin{eqnarray}
A_{CP}'
&=&\frac{2}{|V_{tb}^*V_{ts}|^2|A_t^j|^2(1+\delta_t^j)}
\Big[Im(V_{tb}^*V_{ts}V_{cb}V_{cs}^*)Im(A_t^jA_c^{*j})(1+\delta_c^j)
+Im(V_{tb}^*V_{ts}V_{ub}V_{us}^*)Im(A_t^jA_u^{*j})(1+\delta_u^j)\Big]
\label{83}
\end{eqnarray}
and the error for it is
\begin{eqnarray}
\frac{\Delta A_{CP}}{A_{CP}}
=\frac{A_{CP}'-A_{CP}}{A_{CP}}
\propto
\frac{(\delta_c^j-\delta_t^j)Im(A_c^{j*}/A_t^{j*})+
(\delta_u^j-\delta_t^j)Im(A_u^{j*}/A_t^{j*})}
{(1+\delta_t^j)[Im(A_c^{j*}/A_t^{j*})+Im(A_u^{j*}/A_t^{j*})]}~.
\label{92}
\end{eqnarray}
We can see that the uncertainties can cancel.
We have checked that  numerically
the class(1) error for the CP asymmetry amounts  to few percent
and is small compared to other errors (see below).
\item \underline{Isospin Breaking}

On the other hand, 
we want to show that
the hadronic parameter
uncertainties 
especially from
${\omega_B}$ and ${c}$ dependences
of the isospin breaking effect
can be large even though we take the ratio
as the CP asymmetry.
The decay width
of the neutral and charged decay modes 
with the theoretical error can be written
from eq.(\ref{81}) and (\ref{82}) as
\begin{eqnarray}
\Gamma^0&=&|V_{tb}^*V_{ts}A_t^0(1+\delta_t^0)
+V_{cb}^*V_{cs}A_c^0(1+\delta_c^0)
+V_{ub}^*V_{us}A_u^0(1+\delta_u^0)|^2,\\
\Gamma^+&=&|V_{tb}^*V_{ts}A_t^+(1+\delta_t^+)
+V_{cb}^*V_{cs}A_c^+(1+\delta_c^+)
+V_{ub}^*V_{us}A_u^+(1+\delta_u^+)|^2,
\end{eqnarray}
and the isospin breaking effect is given by
\begin{eqnarray}
\Delta_{0+}'
=\frac{|A_t^0|^2(1+\delta_t^0)^2-|A_t^+|^2
(1+\delta_t^+)^2}{|A_t^0|^2(1+\delta_t^0)^2+|A_t^+|^2
(1+\delta_t^+)^2}~,
\end{eqnarray}
where we neglected all terms except for
those proportional to
${|A_t^j|^2}$ because
${|A_c^j|^2 /|A_t^j|^2 \sim O(10^{-4})}$,
and the CKM factor of the ${|A_u^j|^2}$ is
suppressed as 
${|V_{ub}V_{us}^*/V_{tb}^*V_{ts}|^2 \sim O(\lambda^4)}$.
Then the error can be expressed as
\begin{eqnarray}
\frac{\Delta (\Delta_{0+})}{\Delta_{0+}}
 =\frac{\Delta_{0+}'-\Delta_{0+}}{\Delta_{0+}}
 \approx 
 \frac{4|A_t^0|^2|A_t^+|^2\Big[\delta_t^0-\delta_t^+\Big]}
 {(|A_t^0|^2-|A_t^+|^2)(|A_t^0|^2+|A_t^+|^2)}~.\label{40}
\end{eqnarray}
We can easily imagine that the decay constant
uncertainties are canceled as the direct CP asymmetry.
However
we observe that 
even though ${\delta_t^0-\delta_t^+}$
is small, there exist ${|A_t^0|^2-|A_t^+|^2}$ in the denominator
and it is also small, then the error enhancement can occur.
Variation of ${\omega_B}$ and ${c}$ 
introduces   
${\delta_t^0-\delta_t^+\simeq 0.5\%}$ while
${(|A_t^0|^2-|A_t^+|^2)/(|A_t^0|^2+|A_t^+|^2)\simeq 5\%}$.
This gives about 20\% error for the isospin breaking.
From the above argument, we can see that the
error from ${\omega_B}$
and ${c}$ 
uncertainties
remain somewhat large.
Thus we estimate the class(1) error for the
isospin breaking effect
to be about 20\%. 

\end{itemize}

\item 
Next we want to discuss the class(2) error.
For class(2),
we expect an error coming from the
fact that we used the leading order term in ${\alpha_s(t)}$.
There are also errors coming from neglecting higher order decay amplitudes.
But we have not checked the effect of class(2) errors
as it requires actual computation of higher order amplitudes.
We guess that the error is approximately 15\% in the amplitude.
Then the theoretical errors from class(2)
are 30\% in the branching ratio,
about a few \% in the direct CP asymmetry,
and 20\% in the isospin breaking effect.

\item 
About the class(3) error,
we change the ${\bar{\rho}}$, ${\bar{\eta}}$
parameter in the range ${\bar{\rho}=\rho(1-\lambda^2/2)}$
${=0.20\pm 0.09}$
and ${\bar{\eta}=\eta(1-\lambda^2/2)=0.33\pm 0.05}$ \cite{Eidelman:2004wy},
and numerically estimate how the physical quantities
are affected by the changing of parameters.
The major contributions to the
branching ratio and isospin breaking effects
come from the terms which are proportional to
${V_{tb}^*V_{ts}}$, so they are less sensitive
to the error in ${\bar{\rho}}$, ${\bar{\eta}}$.
On the other hand, direct CP asymmetry
depends on ${Im(V_{tb}^*V_{ts}V_{cb}V_{cs}^*)}$
and ${Im(V_{tb}^*V_{ts}V_{ub}V_{us}^*)}$ 
as in eq.(\ref{83}),
thus the error from the ${\bar{\rho}}$, ${\bar{\eta}}$ 
uncertainties
amounts to about 15\%.

\item  
The class(4) error
comes from the ${u}$ quark loop hadronic uncertainties.
The terms which are proportional to ${V_{ub}^*V_{us}}$
are not very important
to the computation of the branching ratio and isospin breaking
effect, so for these quantities
we can neglect the class(4)
uncertainties.

However
for CP asymmetry,
${c}$ and ${u}$ quark loops
give comparable contributions as seen in eq.(\ref{83}),
thus the ${u}$ quark loop contribution
which is infected with nonperturbative
correction, cannot be neglected.
If we regard the ${u}$ quark loop uncertainty
as about 100\% at the amplitude level for
both real and imaginary part,
the numerical error
for the direct CP asymmetry
amounts to about 75\%.
\end{enumerate}

In summary, we regard the error of the branching ratio,
direct CP asymmetry, and isospin breaking effects
as 50\% (class(1); 40\%, class(2); 30\%), 
75\% (class(4); 75\%), 
and 30\% (class(1); 20\%, class(2); 20\%), respectively.

\end{widetext}

\subsection{Numerical Results}
The numerical results for each decay amplitude ${M_i}$
in the neutral decay (Table\ref{neutral}) and charged decay 
(Table\ref{charged})
in unit of ${10^{-6} \mbox{GeV}^{-2}}$ are as follows. 

The total decay amplitude can be expressed by using these
components as
\begin{eqnarray}
A(B\to K^*\gamma)&=&
M_t+M_c+M_u\nonumber\\
&=&(\epsilon_{\gamma}^*\cdot
\epsilon_{K^*}^*)(M_t^S+M_c^S+M_u^S)\nonumber\\
&+&i\epsilon_{\mu\nu +-}\epsilon_{\gamma}^{*\mu}\epsilon_{K^*}^{*\nu}
(M_t^P+M_c^P+M_u^P),
\end{eqnarray}
\begin{eqnarray*}
M_t&=&M_{7\gamma}+M_{8g}+M_{3\sim 6},\\
M_c&=&M_{1c}+M_{2c}+M_{\psi},  \\
M_u &=& M_{1u+2u}
+M_2+M_{\rho+\omega },
\end{eqnarray*}
where all components include CKM factors.
If we express ${K^*}$ and ${\gamma}$ helicities
as ${\lambda_1,\lambda_2}$,
the combinations which can contribute
to the decay amplitude are 
${A_{\lambda_1,\lambda_2}=A_{+,+},A_{-,-}}$,
if we take into account the fact that
${B}$ meson is spinless and a real photon has
helicities ${\pm 1}$.
Then the total decay width of ${B\to K^*\gamma}$
is given by 
\begin{eqnarray}
\Gamma=\frac{1}{8\pi M_B}\left(
|M_t^S+M_c^S+M_u^S |^2+|M_t^P+M_c^P+M_u^P |^2
\right),\hspace{2mm}
\end{eqnarray}
and the branching ratios for ${B\to K^*\gamma}$
become as follows:
\begin{eqnarray}
Br(B^0\to K^{*0}\gamma)&=&
(5.8\pm 2.9)\times 10^{-5}\label{b0},\\
Br(B^{\pm}\to K^{*\pm}\gamma)&=&
(6.0\pm 3.0)\times 10^{-5}.\label{barb0}
\end{eqnarray}

\begin{figure}
\includegraphics[width=5cm]{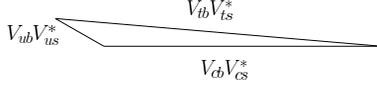}
\caption{CKM unitary triangle}
\label{KM}
\end{figure}

Next we want to extract the direct CP asymmetry.
We take into account 
up to ${O(\lambda^4)}$
about the CKM matrix components,
\begin{eqnarray}
&&\hspace{-5mm}\footnotesize{V_{KM}}=\nonumber\\
&&\hspace{-5mm}
\begin{pmatrix}
\footnotesize{1-\lambda^2/2-\frac{\lambda^4}{8}} &
 \footnotesize{\lambda} &\footnotesize{ A{{\lambda}^3}(\rho -i\eta)}\nonumber\\
\footnotesize{-\lambda} & \footnotesize{1-\lambda^2/2
-\left(1/8+A^2/2\right)
\lambda^4} & \footnotesize{A{{\lambda}^2}}\nonumber\\
\footnotesize{A{{\lambda}^3}(1-\rho-i\eta)}
&\footnotesize{-A\lambda^2 +A\lambda^4\left(1/2
-\rho-i\eta\right)}
&\footnotesize{1-A^2\lambda^4/2}\nonumber
\end{pmatrix}
\\
\end{eqnarray}
and the unitary triangle related to this
decay mode should be crushed (Fig.\ref{KM}).

If we express each amplitudes ${M_{i}}$
as ${\xi_i A_ie^{i\delta_i}}$ where ${\xi_i=V_{ib}^*V_{is}/V_{cb}^*V_{cs}}$,
in order to separate weak phase and strong phase ${\delta_i}$,
the decay amplitudes can be rewritten as
\begin{eqnarray}
&&\hspace{-1.5cm}A(B\to K^*\gamma)=V_{cb}^*V_{cs}\hspace{1mm}[
\xi_t A_t e^{i\delta_t}+ \xi_cA_c e^{i\delta_c}+
\xi_uA_u e^{i\delta_u}],
\\
&&\hspace{-1.5cm}A(\bar B\to \bar{K^*}\gamma)=V_{cb}V_{cs}^*\hspace{1mm}[
\xi_t^* A_t e^{i\delta_t}+ \xi_c^* A_c e^{i\delta_c}+
\xi_u^*A_u e^{i\delta_u}],
\end{eqnarray}
and the direct CP asymmetry can be expressed
as
\begin{eqnarray}
A_{CP}\equiv 
\frac{\Gamma(\bar B \to \bar{K^*}\gamma)-\Gamma(B\to K^*\gamma)}
{\Gamma(\bar B \to \bar{K^*}\gamma)+\Gamma(B\to K^*\gamma)}\equiv
\frac{R_N}{R_D},
\end{eqnarray}
\begin{eqnarray}
R_N&=&\Big[
A_tA_c\sin(\delta_t-\delta_c)\mbox{Im}(V_{tb}V_{ts}^*V_{cb}^*V_{cs})\nonumber\\
&&\hspace{0.5cm}+A_cA_u\sin(\delta_c-\delta_u)\mbox{Im}(V_{cb}V_{cs}^*V_{ub}^*V_{us})
\nonumber\\
&&\hspace{0.5cm}+A_uA_t\sin(\delta_u-\delta_t)\mbox{Im}(V_{ub}V_{us}^*V_{tb}^*V_{ts})
\Big],\\
R_D
 &=&\left(A_t^2|V_{tb}V_{ts}^*|^2+A_c^2|V_{cb}V_{cs}^*|^2+A_u^2|V_{ub}V_{us}^*|^2\right)/2\nonumber\\
&&\hspace{5mm}
+A_tA_c\cos(\delta_t-\delta_c)\mbox{Re}(V_{tb}V_{ts}^*V_{cb}^*V_{cs})\nonumber\\
&&\hspace{5mm}+A_cA_u\cos(\delta_c-\delta_u)\mbox{Re}(V_{cb}V_{cs}^*V_{ub}^*V_{us})\nonumber\\
&&\hspace{5mm}+A_uA_t\cos(\delta_u-\delta_t)\mbox{Re}(V_{ub}V_{us}^*V_{cb}^*V_{cs}),
\end{eqnarray}
then  its' values are
\begin{eqnarray}
A_{CP}(B^0\to K^{*0}\gamma)&=&
-(6.1\pm 4.6)\times 10^{-3},\\
A_{CP}(B^{\pm}\to K^{*\pm}\gamma)&=&
-(5.7\pm 4.3)\times 10^{-3}.
\end{eqnarray}

Finally, we want to estimate the isospin breaking effects
 as eq.(\ref{Isospin}).
This effect is caused by ${O_{8g}}$ (Fig.\ref{O8g}), ${c}$ and ${u}$
loop
contributions (Fig.\ref{cloop}), ${O_1\sim O_{6}}$
annihilation (Figs.\ref{annihilation} and \ref{penguin}), 
and the long distance contributions mediated
${\rho}$ and ${\omega}$ in charged mode (Fig.\ref{Long}).
About the
bremsstrahlung photon contributions emitted through quark lines,
whether the spectator quark is ${u}$ or ${d}$
affects the strength and the sign for the
coupling of photon and quark line,
so they generate the isospin breaking effects.
The most important contributions to the isospin breaking effects
come from QCD penguin ${O_5}$, ${O_6}$ annihilation.
These effects are additive to the dominant contribution ${O_{7\gamma}}$
in both neutral and charged decays (see Tables \ref{neutral} and \ref{charged}).
However its' size are different:
the neutral mode's is larger than charged mode's.
Then the sign of total isospin breaking effects becomes plus
and it's
value is as follows.
\begin{eqnarray}
\Delta_{0+}=+(2.7\pm 0.8)\times 10^{-2}
\end{eqnarray}

\begin{widetext}
\begin{center}
\begin{table}
\begin{tabular}{c|c c c || c c c}
\hline
&&${M_i^S/F^{(0)}}$&&&${M_i^P/F^{(0)}}$&\\
\hline
\hline
${V_{tb}^*V_{ts}}$& 
${M_{7\gamma}^S/F^{(0)}}$&
${M_{8g}^S/F^{(0)}}$&
${M_{3\sim 6}^S/F^{(0)}}$&${M_{7\gamma}^P/F^{(0)}}$&
${M_{8g}^P/F^{(0)}}$&${M_{3\sim 6}^P/F^{(0)}}$\\
\hline
&
-218.67-3.86i&-2.19-0.55i&-11.56-5.70i&218.67+3.86i&2.27+0.59i&11.58+5.63i\\
\hline
${V_{cb}^*V_{cs}}$&
 ${M_{1c}^S/F^{(0)}}$&${M_{2c}^S/F^{(0)}}$&${M_{\psi}^S/F^{(0)}}$
& ${M_{1c}^P/F^{(0)}}$&${M_{2c}^P/F^{(0)}}$&${M_{\psi}^P/F^{(0)}}$\\
\hline
&
-0.29-1.01i&6.42-12.63i &
-13.29&-0.19+1.27i&-4.81+8.23i&
15.09\\
\hline
${V_{ub}^*V_{us}}$& 
${M_{1u+2u}^S/F^{(0)}}$&${M_{2}^S/F^{(0)}}$
&${M_{\rho+\omega}^S/F^{(0)}}$
& ${M_{1u+2u}^P/F^{(0)}}$&${M_{2}^P/F^{(0)}}$&
${M_{\rho +\omega}^P/F^{(0)}}$\\
\hline
&-0.63+0.22i&0&-0.03-0.06i&
0.67-0.18i&0& 0.03+0.07i\\
\hline
\end{tabular}
\caption{${B^0\to K^{*0}\gamma}$\hspace{1mm} at ${\bar{\rho}=0.20}$,
 ${\bar{\eta}=0.33}$, ${\omega_B=0.40 \mbox{GeV}}$.}
\label{neutral}
\end{table}

\begin{table}
\begin{tabular}{c| c c c|| c c c}
\hline
&&${M_i^S/F^{(0)}}$&&&${M_i^P/F^{(0)}}$&\\
\hline
\hline
${V_{tb}^*V_{ts}}$& ${M_{7\gamma}^S/F^{(0)}}$&
${M_{8g}^S/F^{(0)}}$&
${M_{3\sim 6}^S/F^{(0)}}$&${M_{7\gamma}^P/F^{(0)}}$&
${M_{8g}^P/F^{(0)}}$&${M_{3\sim 6}^P/F^{(0)}}$\\
\hline
&-218.67-3.86i&-4.89-0.10i&-2.47+0.37i&218.67+3.86i&4.83-0.82i&2.86+0.14i\\
\hline
${V_{cb}^*V_{cs}}$&
 ${M_{1c}^S/F^{(0)}}$& ${M_{2c}^S/F^{(0)}}$&${M_{\psi}^S/F^{(0)}}$
& ${M_{1c}^P/F^{(0)}}$& ${M_{2c}^P/F^{(0)}}$&${M_{\psi}^P/F^{(0)}}$\\
\hline
&-0.66+2.15i&6.42-12.63i&
-13.29&1.39-2.60i& -4.81+8.23i&
15.09\\
\hline
${V_{ub}^*V_{us}}$& ${M_{1u+2u}^S/F^{(0)}}$&${M_{2}^S/F^{(0)}}$
&${M_{\rho+\omega}^S/F^{(0)}}$
& ${M_{1u+2u}^P/F^{(0)}}$&${M_{2}^P/F^{(0)}}$&
${M_{\rho +\omega}^P/F^{(0)}}$\\
\hline
&-0.75+0.51i&0.35+1.01i&-0.04+0.05i&
0.79-0.18i&-0.75+1.16i&0.05-0.05i\\
\hline
\end{tabular}
\caption{${B^{+}\to K^{*+}\gamma}$ \hspace{1mm} at
 ${\bar{\rho}=0.20}$, ${\bar{\eta}=0.33}$, ${\omega_B=0.40 \mbox{GeV}}$.}
\label{charged}
\end{table}
\end{center}
\end{widetext}

\section{Conclusion}
\label{Conclusion}
In this paper, we calculated
the branching ratio, direct CP asymmetry, and isospin breaking effect
within the standard model using the pQCD approach.
It is useful to compare our results with those
existing in the literature.
The 
decay amplitude can be obtained from the transition form factor
\begin{eqnarray}
&&\hspace{-1cm}\langle K^*(P_2,\epsilon_{K^*})\mid iq^{\nu}\bar s  \sigma_{\mu\nu}b 
\mid B(P_1)\rangle \nonumber\\
&&\hspace{2cm}
=-iT_1^{K^*}(0)\epsilon_{\mu\alpha\beta\rho}\epsilon_{K^*}^{\alpha}P^{\beta}
q^{\rho}
\end{eqnarray}
where ${P=P_1+P_2}$, ${q=P_1-P_2}$.
Within the framework of pQCD, we obtain the value of the
${B\to K^*}$ transition form factor as 
${T_1^{K^*}(0)=0.23\pm 0.06}$.
The result can be
compared with the ones extracted by 
another estimation.
In the QCD Factorization,
${T_1^{K^*}(0)=0.27\pm 0.04}$
\cite{Beneke:2001at}
and an updated phenomenological estimate of this
quantity 
with the light-cone distribution amplitudes for the ${K^*}$ meson is
${T_1^{K^*}(0)=0.27\pm 0.02}$ \cite{Ali:2004hn}.
While the central value in the updated result is the same as before, 
the error is reduced by a factor of 2.
In the light-cone QCD sum rule
${T_1^{K^*}(0)=0.38 \pm 0.06}$ \cite{Ball:1998kk},
the lattice
QCD simulation ${T_1^{K^*}(0)=0.32^{+0.04}_{-0.02}}$ 
\cite{DelDebbio:1997kr} and ${T_1^{K^*}(0)=0.25^{+0.05}_{-0.02}}$
\cite{Becirevic}, 
and the covariant light-front approach 
${T_1^{K^*}(0)=0.24}$ \cite{Cheng:2004yj}. 
There are several estimates of the branching ratio
by using the value of ${T_1^{K^*}(0)=0.38 \pm 0.06}$ 
extracted from light-cone QCD
sum rule.
Comparing the results with experiments,
this value of the form factor over estimates the branching ratios
\cite{Ali:2001ez,Bosch:2002bw,Bosch:2001gv}.
Also, it should be noted that ${T_1^{K^*}(q^2)}$ and other related
form factors have been computed in the frame work of pQCD \cite{Chen:2002bq}.
They obtained the central value as ${T_1^{K^*}(0)=0.315}$.
The difference between our results and their's is
the ${K^*}$ meson wave function. We take the new
${K^*}$ wave function parameters
computed in Ref.\cite{Ball:2003sc}.

Note that we have also included the long distance contributions.
If we neglect them, the branching ratios become
${Br(B^0 \to K^{*0}\gamma)=
(5.2\pm 2.6)\times 10^{-5}}$ and
${Br(B^{\pm} \to K^{*\pm}\gamma)=
(5.3\pm 2.7)\times 10^{-5}}$
to be pared with results shown in Eq.(\ref{b0}) \hspace{-1mm}and \hspace{-1mm}(\ref{barb0}).
The ${B\to K^*\psi\to K^*\gamma}$
contribution to the total decay width amounts to
about ${12 \%}$ and also it works 
additive to the branching ratios.
We also emphasize that
we can calculate the annihilation contributions
with the pQCD approach,
and these contribute to the total decay width which amount to
about ${2\sim 10 \%}$.

This analysis predicts less than 1\% direct CP asymmetry within the standard
model.
If we neglect the long distance contributions,
the asymmetries become ${A_{CP}(B^0\to K^{*0}\gamma)=
-(6.7\pm 5.0)\times 10^{-3}}$
and
${A_{CP}(B^{\pm}\to K^{*\pm}\gamma)=
-(7.2\pm 5.4)\times
10^{-3}}$,
and as to isospin breaking effect like 
${\Delta_{0+}=+(2.6\pm 0.8)\times 10^{-2}}$.
The long distance contributions do not seem to 
affect to these asymmetries.

The branching ratio of the neutral decay
is similar to that of the charged decay, in spite of the
difference of the lifetime between them.
This effect is mainly caused by the 4-quark penguin operators ${O_5}$,
${O_6}$.
If we neglect these contributions, the isospin breaking is 
${\Delta_{0+}=-(1.2\pm 0.4)\times 10^{-2}}$,
so we can see that they generate about 4\% isospin breaking effect.
This result is similar to the conclusion of Ref.\cite{Kagan:2001zk}.

${B\to K^*\gamma}$ decay,
as we have first mentioned,
is an attractive decay mode
to test the standard model and search for new physics.
In order to look for the new physics,
we have to reduce the experimental errors.
The error to the 
direct CP asymmetry 
must get smaller than ${1\%}$.
That is to say, we need at least 20 times more data.
This is not possible without the super B factory.

\begin{acknowledgements}
We acknowledge useful discussion with
the pQCD group members, especially
Hsiang-nan Li,
AIS acknowledges support from the Japan Society for the Promotion of Science,
Japan-US collaboration program, and a grant from Ministry of Education, Culture, Sports,
Science and Technology of Japan.

\end{acknowledgements}

\appendix
\section{Brief review of pQCD}
\label{pQCD}
\subsection{Divergences in perturbative diagrams }
\begin{figure}
\includegraphics[width=3.8cm]{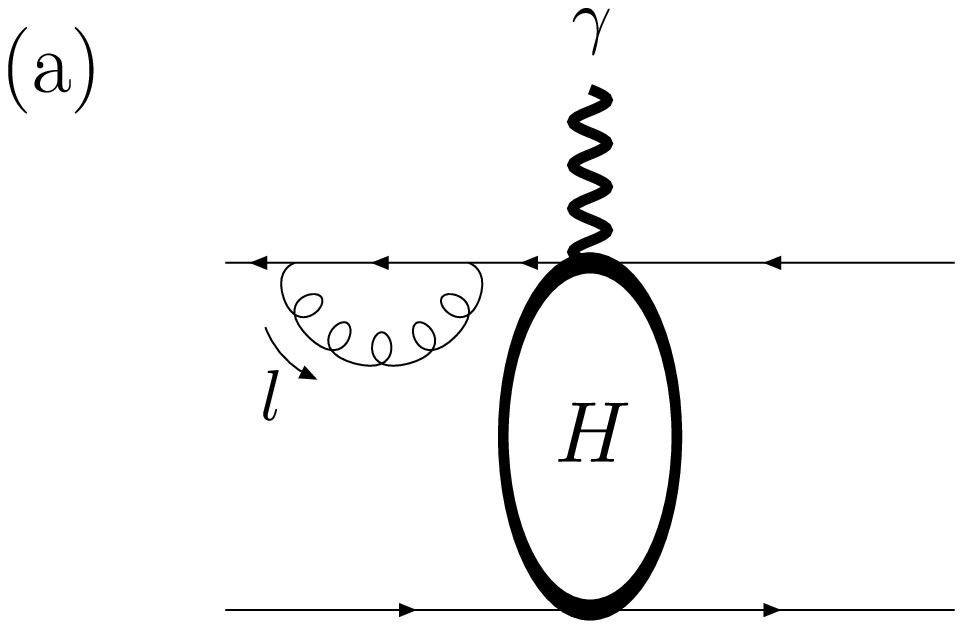}
\hspace{3mm}
\includegraphics[width=3.8cm]{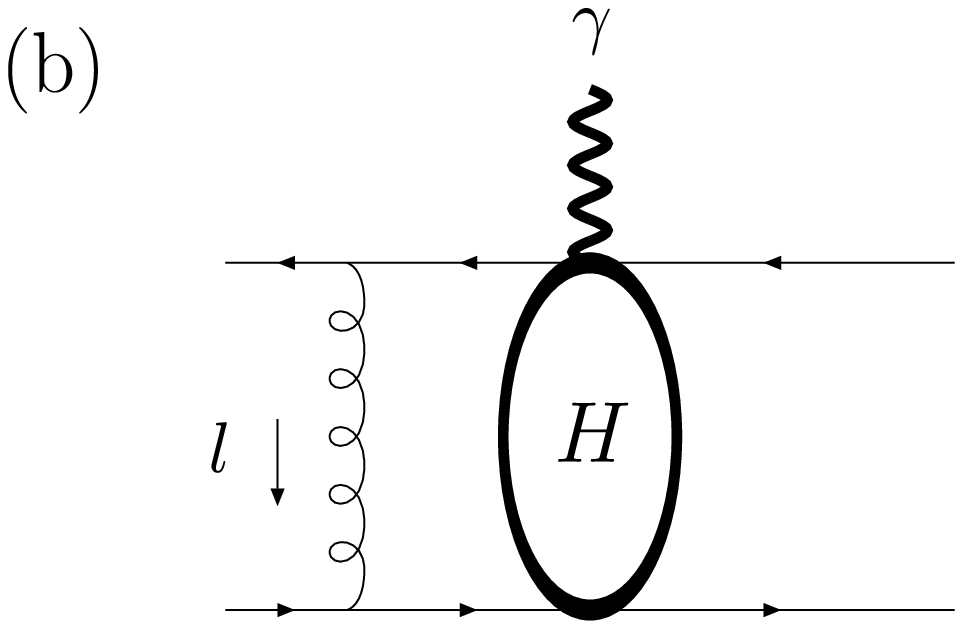}
\vspace{5mm}

\includegraphics[width=3.8cm]{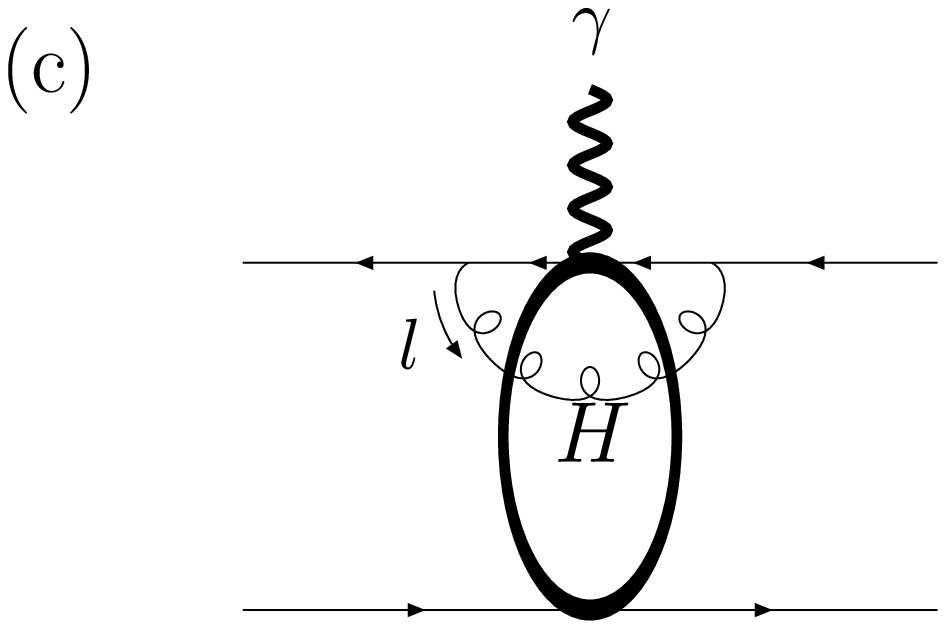}
\hspace{3mm}
\includegraphics[width=3.8cm]{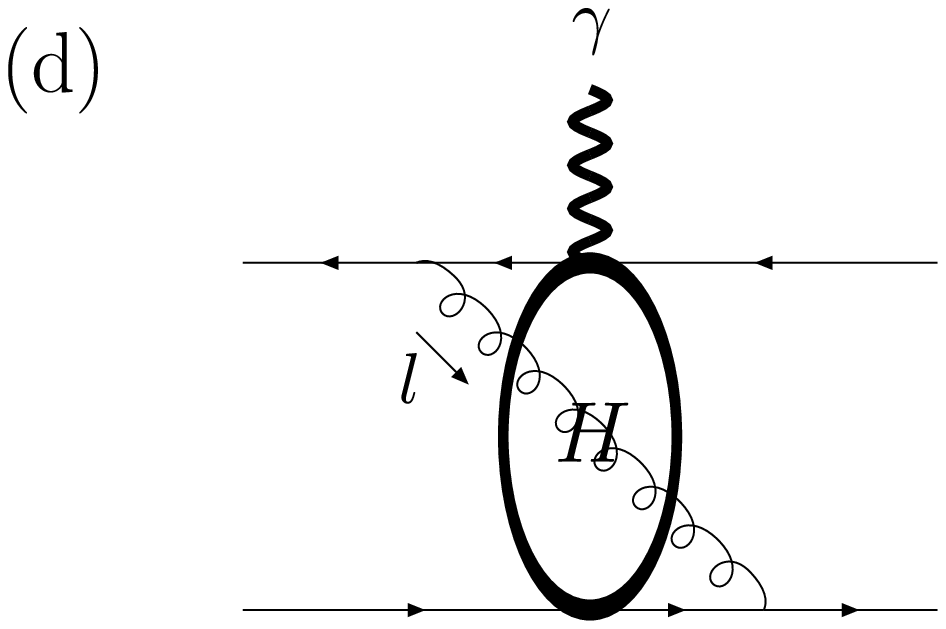}
\caption{${O(\alpha_{s})}$ corrections to the hard scattering ${H}$.}
\label{alpha}
\end{figure}

Here we want to review the ${k_T}$ factorization \cite{Nagashima:2002ia}.
At higher order, infinitely many gluon exchanges must be considered.
In order to understand the factorization procedure, we refer to
the diagrams of Fig.\ref{alpha}.

They describe the ${O(\alpha_s)}$ radiative corrections
to the hard scattering process ${H}$.
In general, individual higher order diagrams have two types 
of infrared divergences: soft and collinear. 
Soft divergence comes from the region of a
loop momentum where all it's momentum components in the
light-cone coordinate vanish:
\begin{eqnarray}
l^{\mu}=(l^+,l^-,\vec{l}_T)=(\Lambda,\Lambda,\vec{\Lambda}).
\end{eqnarray}
Collinear divergence originates
from the gluon momentum region which is parallel
to the massless quark momentum,
\begin{eqnarray}
l^{\mu}=(l^+,l^-,\vec{l}_T)
\sim (M_B,\bar{\Lambda}^2/M_B,\vec{\Lambda}).
\end{eqnarray}
In both cases, the loop integration correspond to 
${\int d^4 l \hspace{1mm}1/l^4 \sim \log{\Lambda}}$,
so logarithmic divergences are generated.
It has been shown order by order in perturbation theory
that these divergences can be separated from
hard kernel 
and absorbed into meson wave functions using eikonal approximation
\cite{Li:1994iu}.

Furthermore, there are also double logarithm divergences in
Fig.\ref{alpha}(a)
and \ref{alpha}(b)
when soft and collinear momentum overlap.
These large double logarithm can be summed
by using renormalization group equation.
This factor is called the Sudakov factor
and also factorized into the definition of 
meson wave function 
\cite{Collins:1981uk,Botts:kf,Li:1992nu}.
The explicit expression for Sudakov factor
is given by \cite{Botts:kf} (see Appendix \ref{Appendix. A}).

There are also
ultraviolet divergences,
and also 
another type of double logarithm
which comes from the loop correction
for the weak decay vertex correction.
These double logarithm can also be factored out from
hard part and grouped into the quark jet function.
These double logarithms also should be resumed
as the threshold factor \cite{Li:2001ay,Kurimoto:2001zj}.
This factor decreases faster than any other power of ${x}$ as 
${x\rightarrow 0}$, so
it removes the endpoint singularity.
Thus we can factor out the
Sudakov factor, the threshold factor, and the ultraviolet divergences
from hard part and grouped into meson wave function
(Appendix \ref{Appendix. A}).
Then the redefinition of wave functions 
including these loop corrections get
factorization energy scale dependence ${t}$.

Thus the amplitude can be factorized into a perturbative part
including a hard gluon exchange, and
a nonperturbative part characterized by the meson distribution amplitudes.
Then the total decay amplitude 
can be expressed as the convolution:
\begin{eqnarray}
&&\hspace{-1cm}\int^1_0
 dx_1 dx_2 \int^{1/\Lambda}_0 d^2b_1 d^2b_2\hspace{1mm}C(t)\otimes
\Phi_{K^*}(x_2,b_2,t)\nonumber\\
&&\hspace{10mm}
\otimes H(x_1,x_2,b_1,b_2,t)\otimes\Phi_B(x_1,b_1,t),
\end{eqnarray}
here ${\Phi_{K^*}(x_2,b_2,t)}$, ${\Phi_B(x_1,b_1,t)}$ are
meson distribution amplitudes that
contain the soft divergences which come from quantum correction 
and 
${H(x_1,x_2,b_1,b_2,t)}$
is the hard kernel including finite piece of quantum correction,
where ${b_1}$, ${b_2}$ are the conjugate variables
to transverse momentum, and 
${x_1}$, ${x_2}$ are the momentum fractions
of spectator quarks.

\subsection{Physical interpretation of Sudakov factor}
\begin{figure}
\begin{minipage}[c]{7cm}
\hspace{-6cm}(a)\\
\includegraphics[width=5.5cm]{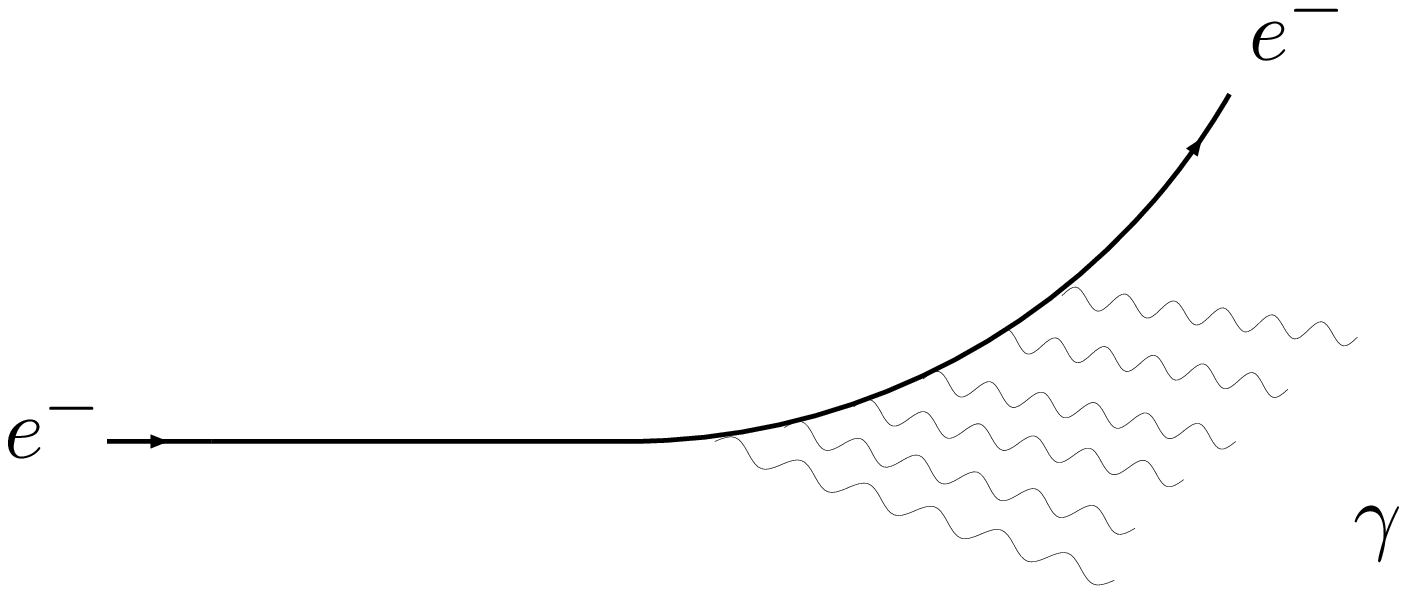}
\end{minipage}
\hspace{1cm}
\begin{minipage}[c]{8cm}
\hspace{-6cm}(b)\\
\includegraphics[width=6cm]{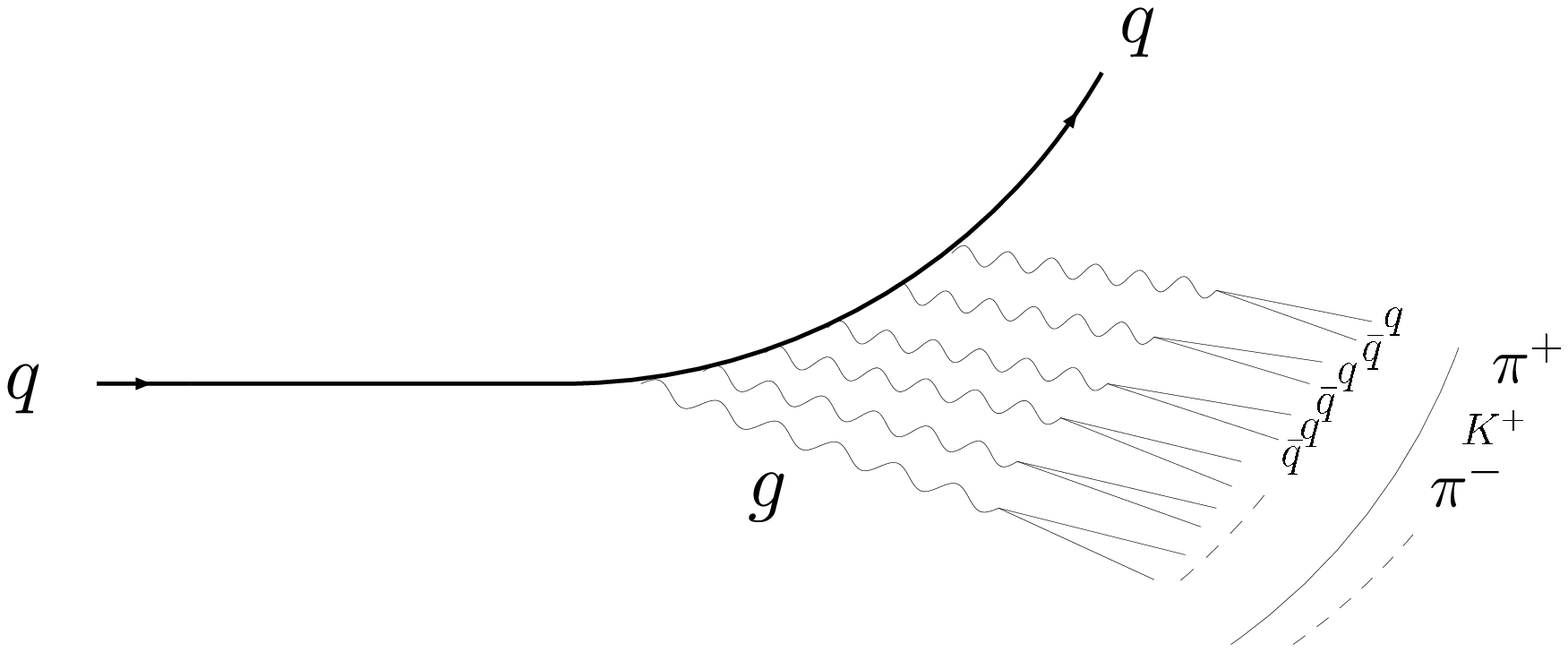}
\end{minipage}
\caption{An electron which is scattered by the electromagnetic
 interaction (a) is observed with many soft photons.
Similarly, a quark which is scattered by the strong interaction (b)
is not observed as a single gluon: accompanied by many soft gluons,
and they form hadron jets.}
\label{photon}
\end{figure}

\begin{figure}
\includegraphics[width=3cm]
{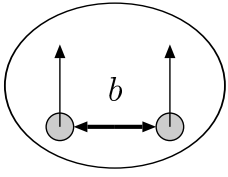}
\caption{${b}$ is the transverse interval between the quark
and antiquark pair in the ${B}$ meson.}
\label{shielding}
\end{figure}

\begin{figure}
\begin{center}
\includegraphics[width=8cm]{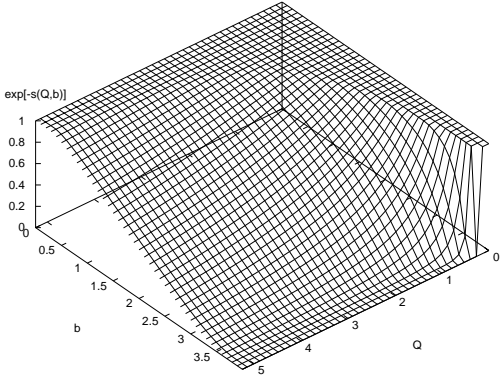}
\caption{The dependence of the Sudakov factor ${\exp[-s(Q,b)]}$
on ${Q}$ and ${b}$ where ${Q}$ is the ${b}$ quark momentum,
and ${b}$ is the interval between quarks
which form hadrons. It is clear that the large ${b}$ and ${Q}$
region is
suppressed.}
\label{Sudacov}
\end{center}
\end{figure}

In order to understand the Sudakov factor physically,
first we consider QED.
When a charged particle is accelerated,
infinitely many photons must be emitted by the bremsstrahlung 
(Fig.\ref{photon}(a)).
A similar phenomenon occurs when a quark 
is accelerated:
infinitely many gluons must be emitted.
According to the feature of strong interaction,
gluons cannot exist freely,
so hadronic jet is produced.
Then we observe many hadrons
in the end if gluonic bremsstrahlung occurs.
Thus the amplitude for an exclusive decay
${B\to K^*\gamma}$ is proportional to the probability that
no bremsstrahlung gluon is emitted.
This is the Sudakov factor and it is depicted in Fig.\ref{Sudacov}.
As seen in Fig.\ref{Sudacov}, the Sudakov factor is large for small
${b}$
and ${Q}$. 
Large ${b}$ implies that the quark and antiquark pair is
separated,
which  in turn implies less color shielding (see Fig.\ref{shielding}).
Similar absence of shielding occurs when ${b}$ quark carries
most of the momentum while the momentum fraction of spectator quark 
${x}$ in the ${B}$ meson is small.

Then the Sudakov factor suppresses the long distance contributions 
for the decay process and gives the effective cutoff
about the transverse direction \cite{Li:1992nu,Collins:ta}.
In short,  the Sudakov factor corresponds to the probability
for emitting no photons.
According to this factor,
the property of short distance is guaranteed.

\section{Some functions}
\label{Appendix. A}
The expressions for some functions are presented in this appendix.
In our numerical calculation, we
use the leading order ${\alpha_s}$ formula. 
\begin{eqnarray}
\hspace{-5mm}\alpha_s(\mu)=\frac{2\pi}{\beta_0\ln(\mu/\Lambda_{n_f})}_,\hspace{10mm}
\beta_0 &=&\frac{33-2n_f}{3}\hspace{3mm}
\end{eqnarray}
The explicit expression for 
Sudakov factor ${s(t,b)}$ is given by \cite{Botts:kf}
\begin{eqnarray}
s(t,b)=\int^t_{1/b}\frac{d\mu}{\mu}
\left[\ln{\left(\frac{t}{\mu}\right)}
A(\alpha_s(\mu))+B(\alpha_s(\mu))
\right],
\end{eqnarray}
\begin{eqnarray}
A&=&C_F \frac{\alpha_s}{\pi}+
\left(\frac{\alpha_s}{\pi}\right)^2\nonumber\\
&&\times 
\left[\frac{67}{9}-\frac{{\pi}^2}{3}-\frac{10}{27}n_f +
\frac{2}{3}\beta_0\ln \left(\frac{e^{\gamma}_E}{2}\right)
\right],\\
B&=&\frac{2}{3}\frac{\alpha_s}{\pi}\ln\left(\frac{e^{2\gamma_E}-1}{2}\right),\\
\nonumber
\end{eqnarray}
where ${\gamma_E =0.5722}$ is Euler constant and ${C_F=4/3}$ is color factor.
The meson wave function including summation factor
has energy dependence
\begin{eqnarray}
\phi_B(x_1,b_1,t)&=&\phi_B(x_1,b_1)\exp{[-S_B(t)]},\\
\phi_{K^*}(x_2,t)&=&\phi_{K^*}(x_2)\exp{[-S_{K^*}(t)]},
\end{eqnarray}
and the total functions including the Sudakov factor
and the ultraviolet divergences are
\begin{eqnarray}
S_B(t)&=&s(x_1P_1^-,b_1)+2\int^t_{1/b_1}\frac{d\bar{\mu}}{\bar{\mu}}\gamma
(\alpha_s(\bar{\mu})),\\
S_{K^{\ast}}(t)&=&s(x_2P_2^-,b_2)+s((1-x_2)P_2^-,b_2)\nonumber\\
&&\hspace{1cm}+2\int^t_{1/b_2}
\frac{d\bar{\mu}}{\bar{\mu}}\gamma(\alpha_s(\bar{\mu})).
\end{eqnarray}
Threshold factor is expressed as
below
\cite{Li:2001ay,Keum:2000wi},
and  we take the value ${c=0.4}$.
\begin{eqnarray}
S_t(x)=\frac{2^{1+2c}\Gamma(3/2+c)}{\sqrt{\pi}\Gamma(1+c)}
[x(1-x)]^c
\end{eqnarray}
\section{Wave functions}
\label{Appendix. B}
For the ${B}$ meson wave function, we adopt the model
\begin{eqnarray}
\Phi_B(P_1)&=&\frac{1}{\sqrt{2N_c}}(\Slash{P}_1+M_B)\gamma^5\phi_B(k_1),\\
\phi_B(x_1,b_1)
&=&\int dk_1^- d^2k_{1\perp}e^{i\vec{k}_{1\perp}\cdot
 \vec{b}}\phi_B(k_1)\nonumber\\
&&\hspace{-22mm}=N_B
 x_1^2(1-x_1)^2\exp{\left[-\frac{1}{2}\left(\frac{x_1M_B}{\omega_B}\right)^2-\frac{b_1^2\omega_B^2}{2}\right]},
\end{eqnarray}
with th shape parameter ${\omega_B=(0.40\pm 0.04)}$GeV.
The normalization constant ${N_B}$ is fixed by the decay
constant ${f_B}$
\begin{eqnarray}
\int^1_0 dx \phi_B(x,b=0)=\frac{f_B}{2\sqrt{2N_c}}_,
\end{eqnarray}
where ${N_c}$ is the color number. 

We use the vector meson wave functions
determined by the light-cone QCD sum rule
\cite{Ball:1998ff,Ball:2003sc}.
We choose the vector meson momentum ${P}$ 
moving in  the ``-'' direction along the ${z}$
axis
with
${P^2=M_{V}^2}$, and the polarization vectors ${\epsilon_L}$,
${\epsilon_T}$ are defined as
\begin{eqnarray}
\epsilon_{L}=(0,1,\vec{0}),\hspace{6mm}
\epsilon_{T}=\left(0,0,\frac{1}{\sqrt{2}}(\pm 1,-i)\right),
\end{eqnarray}
and ${\epsilon_T}$
satisfies the gauge invariant condition ${P\cdot \epsilon_T =0}$.
The nonlocal matrix elements sandwiched between the vacuum and the
${K^*}$ meson state can be expressed as follows,
\begin{eqnarray}
&&
\langle K^{\ast -}(P) \mid\bar{s}(z)Iu(0)\mid 0\rangle =\frac{1}{2N_c} 
f_{K^{\ast}}^T\nonumber\\
&&\hspace{2cm}\times \frac{\epsilon_L\cdot z}{p\cdot z}M_{K^{\ast}}^2
\int_0^1 dx e^{ix P \cdot z} 
\frac{\partial}{\partial x}h_{\parallel}^{(s)}(x)\nonumber\\
\end{eqnarray}
\begin{eqnarray}
&&\langle K^{\ast -}(P) \mid\bar{s}(z)\gamma_{\mu}u(0)\mid 0\rangle
=\frac{f_{K^{\ast}}}{N_c}M_{K^{\ast}}
 \Big[
P_{\mu}\frac{\epsilon_L\cdot z }
{P\cdot z}\nonumber\\
&&\hspace{7mm}\times\int_0^1 dx e^{ix P \cdot z}\phi_{\parallel}(x)+\epsilon_{T
 \mu} \int_0^1 dx e^{ix P \cdot z}g_{\perp}^{(v)}(x)\Big]\nonumber\\
\end{eqnarray}
\begin{eqnarray}
&&\langle K^{\ast -}(P) \mid\bar{s}(z)\gamma_{5}\gamma_{\mu} u(0)\mid 
0\rangle =-\frac{i}{4N_c}f_{K^{\ast}}\nonumber\\
&&\hspace{5mm}\times \frac{M_{K^{\ast}}}{P\cdot z}
\epsilon_{\mu\nu\rho\sigma}\hspace{1mm}\epsilon_{T}^{\nu}P^{\rho}
z^{\sigma} \int_0^1 dxe^{ixP \cdot z} \frac{\partial}{\partial x}g_{\perp}^{(a)}(x)
\nonumber\\
\end{eqnarray}
\begin{eqnarray}
&&\langle K^{\ast -}(P) \mid\bar{s}(z)\sigma_{\mu\nu} u(0)\mid 0\rangle
=-i\frac{f_{K^{\ast}}^T }{N_c}\nonumber\\
&&\times \Big[(\epsilon_{T\mu}P_{\nu}-\epsilon_{T\nu}
P_{\mu})\int_0^1 dx e^{ixP \cdot z}\phi_{\perp}(x)\nonumber\\
&&+(P_{\mu}z_{\nu}-P_{\nu}z_{\mu})\frac{\epsilon_L\cdot z}
{(P\cdot z)^2}M_{K^{\ast}}^2 \int_0^1 dx e^{ixP \cdot z}
h_{\parallel}^{(t)}(x)
\Big]\nonumber\\
\end{eqnarray}
where we neglect the terms proportional to ${r_{K^*}^2}$ (twist-4)
and the terms ${(m_u+m_s)/M_{K^*}}$.
Then the ${K^*}$ meson distribution amplitudes up to twist-3 are
\begin{eqnarray}
\Phi_{K^*}^L(P,\epsilon_{ L})
&=&\frac{1}{\sqrt{2N_c}}\int_0^1 dx e^{ixP\cdot z}\{M_{K^{\ast}}[\Slash{\epsilon}_{L}]\phi_{K^{\ast}}(x)\nonumber\\
&&
+[\Slash{\epsilon}_{
L}\hspace{1mm}\Slash{P}]\phi_{K^{\ast}}^t(x)
+M_{K^{\ast}}[I]\phi_{K^{\ast}}^s(x)\},\nonumber\\
\\
\Phi_{K^*}^T(P,\epsilon_{T})
&=&\frac{1}{\sqrt{2N_c}}\int_0^1 dx e^{ixP\cdot z}
\{
M_{K^{\ast}}[\Slash{\epsilon}_{T}]\phi_{K^{\ast}}^v(x)\nonumber\\
&&\hspace{-18mm}
+[\Slash{\epsilon}_{
 T}\hspace{1mm}\Slash{P}]\phi_{K^{\ast}}^T(x)\hspace{1mm}
+\frac{M_{K^{\ast}}}{P\cdot z}i\epsilon_{\mu\nu\rho\sigma}
 [\gamma^{\mu}\gamma^5]
\epsilon_T^{\nu}P^{\rho}z^{\sigma}\phi_{K^{\ast}}^a(x)\}\nonumber\\
\end{eqnarray}
\begin{eqnarray}
&&\hspace{-5mm}\phi_{K^{\ast}}(x)=\frac{f_{K^{\ast}}}{2\sqrt{2N_c}}\phi_{\parallel},
\hspace{3mm}\phi_{K^{\ast}}^t(x)=\frac{f_{K^{\ast}}^T}{2\sqrt{2N_c}}h_{\parallel}^{(t)},\nonumber\\
&&\hspace{-5mm}\phi_{K^{\ast}}^s(x)=\frac{f_{K^{\ast}}^T}{4\sqrt{2N_c}}
\frac{d}{dx}h_{\parallel}^{(s)},\hspace{3mm}
\phi_{K^{\ast}}^T(x)=\frac{f_{K^{\ast}}^T}{2\sqrt{2N_c}}\phi_{\perp}
,\nonumber\\
&&\hspace{-5mm}\phi_{K^{\ast}}^v(x)=\frac{f_{K^{\ast}}}{2\sqrt{2N_c}}g_{\perp}^{(v)},
\hspace{3mm}\phi_{K^{\ast}}^a(x)=\frac{f_{K^{\ast}}}{8\sqrt{2N_c}}
\frac{d}{dx}g_{\perp}^{(a)},
\nonumber\\
\end{eqnarray}
where we use ${\epsilon_{0123}=1}$ and
set the normalization condition
about ${ \phi_i =\{\phi_{\parallel},\phi_{\perp},g_{\perp}^{(v)}
,g_{\perp}^{(a)},h_{\parallel}^{(t)},h_{\parallel}^{(s)}\}}$ 
as
\begin{eqnarray}
\int_0^1dx \phi_i(x)=1.
\end{eqnarray}

\begin{center}
\begin{table}
\begin{tabular}{c||c c }
\hline
${V}$ & ${K^*}$ & ${\rho}$\\
\hline
${f_{V}}$[MeV] & ${226\pm 28}$ & ${198\pm 7}$\\
\hline
${f_{V}^T}$[MeV] & ${185\pm 10}$ & ${160\pm 10}$\\
\hline
${a_{1}^{\parallel}}$[MeV] & ${-0.4\pm 0.2}$ & 0\\
\hline
${a_2^{\parallel}}$[MeV] & ${0.09\pm 0.05}$& ${0.18\pm 0.10}$\\
\hline
${a_1^{\perp}}$[MeV] & ${-0.34\pm 0.18}$ & 0\\
\hline
${a_2^{\perp}}$[MeV] & ${0.13\pm 0.09}$ & ${0.2\pm 0.1}$\\
\hline
${\delta_+}$ & 0.24 & 0\\
\hline
${\delta_-}$ & -0.24 & 0\\
\hline
${\tilde{\delta}_+}$ & ${0.16}$ & 0\\
\hline
${\tilde{\delta}_-}$ & -0.16 & 0\\
\hline
${\zeta_3^A}$ & 0.032 & 0.032\\
\hline
${\zeta_3^V}$&0.013 & 0.013\\
\hline
${\zeta_3^T}$& 0.024 & 0.024\\
\hline
${\omega_{1,0}^A}$ & -2.1 & -2.1\\
\hline
\end{tabular}
\caption{Some parameter quantities.}
\label{P}
\end{table}
\end{center}

\begin{eqnarray*}
\phi_{\parallel}(x)&=&6x(1-x)\left[1+3a_1^{\parallel}x_i
 +\frac{3}{2}a_2^{\parallel}(5x_i^2 -1)
\right]
\end{eqnarray*}
\begin{eqnarray*}
\phi_{\perp}(x)&=&6x(1-x)\left[1+3a_1^{\perp}x_i
 +\frac{3}{2}a_2^{\perp}(5x_i^2 -1)
\right]
\end{eqnarray*}
\begin{eqnarray*}
h_{\parallel}^{(s)}(x)&=&6x(1-x)\left[1+a_1^{\perp}x_i+
\left(\frac{1}{4}a_2^{\perp}
+\frac{35}{6}\xi_3^T\right)(5x_i^2-1)\right]\nonumber\\
&&+3\delta_+\left[3x(1-x)+x\ln x +(1-x)\ln (1-x)\right]\nonumber\\
&&+3\delta_-
\left[x\ln x -(1-x)\ln(1-x)\right]
\end{eqnarray*}
\begin{eqnarray*}
h_{\parallel}^{(t)}(x)&=&3x_i^2 +\frac{3}{2}a_1^{\perp}x_i(3x_i^2-1)
+\frac{3}{2}a_2^{\perp}x_i^2(5x_i^2-3)\nonumber\\
&&+\frac{35}{4}\zeta_{3}^T (3-30x_i^2 +35x_i^4)
+\frac{3}{2}\delta_+\left[1+x_i\ln\left(\frac{x}{1-x}\right)\right]\nonumber\\
&&+\frac{3}{2}\delta_- {x_i}\left[2+\ln x +\ln (1-x)\right]
\end{eqnarray*}
\begin{eqnarray*}
g_{\perp}^{(a)}(x)&=&6x(1-x)\Big[
1+a_1^{\parallel}\xi +\{\frac{1}{4}a_2^{\parallel}+\frac{5}{3}
\zeta_3^A\left(1-\frac{3}{16}\omega_{1,0}^A\right)\nonumber\\
&&+\frac{35}{4}\zeta_3^V\}
(5x_i^2-1)\Big]+6\tilde{\delta}_+\Big[3x(1-x)+x\ln x \nonumber\\
&&+(1-x)\ln (1-x)\Big]
+6\tilde{\delta}_-\left[x\ln x -(1-x)\ln(1-x)\right]\nonumber
\end{eqnarray*}
\begin{eqnarray*}
g_{\perp}^{(v)}(x)&=&\frac{3}{4}(1+x_i^2)+a_1^{\parallel}\frac{3}{2}x_i^3
+\left(\frac{3}{7}a_2^{\parallel}+5\zeta_3^A\right)(3x_i^2-1)\nonumber\\
&&+\left(\frac{9}{112}a_2^{\parallel}+\frac{105}{16}\xi_3^V -\frac{15}{64}
\xi_3^A \omega_{1,0}^A\right)(3-30x_i^2 +35x_i^4)\nonumber\\
&&+\frac{3}{2}\tilde{\delta}_+\left[2+\ln x+\ln(1-x)\right]
\nonumber\\&&+\frac{3}{2}
\tilde{\delta}_-\left[2x_i +\ln(1-x) -\ln x\right]\nonumber
\end{eqnarray*}

Here ${x_i=1-2x}$, and 
the expressions about ${\rho}$ and ${\omega}$
meson wave functions are the same as above
with
the values of parameters as
follows evaluated at ${\mu=}$1GeV (Tab.\ref{P}). 
Since ${\rho/\omega}$ states are
${(|\bar{u}u\rangle \mp |\bar{d}d\rangle)/\sqrt{2}}$,
the ${\bar{q}q}$ distribution where ${q=u}$ or ${d}$
can be taken to the same for ${|\bar{u}u\rangle}$
and ${|\bar{d}d\rangle}$ using isospin symmetry.


\end{document}